\documentclass[prd,preprint,aps,amsmath,amssymb,showpacs,superscriptaddress,nofootinbib]{revtex4-2}

\usepackage{xcolor}

\usepackage{graphicx} 
\usepackage{tabularx}
\usepackage{dcolumn}
\usepackage{bm}
\usepackage{hyperref}
\usepackage{color}
\usepackage[normalem]{ulem}
\usepackage{comment}
\usepackage{diagbox}
\usepackage{amsmath}
\renewcommand\[{\begin{equation}} 
\renewcommand\]{\end{equation}}
\usepackage{slashed}
\usepackage{mathtools}

\usepackage{subcaption} 
\usepackage{caption} 
\usepackage{float} 
\usepackage{subcaption}
\usepackage{ragged2e}

\newcommand{\MR}[1]{\textbf{\textcolor{red}{[MR: #1]}}}
\usepackage{ulem}
\renewcommand\[{\begin{equation}} 
\renewcommand\]{\end{equation}}

\begin{abstract}
    Primordial black holes (PBHs) formed during first-order phase transitions provide a powerful link between the early-universe microphysics and observable signatures today, including dark matter and gravitational waves. In this work we develop a unified description of PBH formation based on the Israel junction conditions, which capture collapse dynamics without relying on conventional overdensity or pressure-balance arguments. As a first application, we show that exotic objects such as Fermi-balls can collapse into PBHs even when most of the vacuum energy is trapped in solitonic cores, leading to a different gravitational-wave signal relative to vacuum-only scenarios. As a second application, we study multiple phase transitions in a hidden sector, which generate correlated gravitational-wave spectra and PBH abundances across transitions. Our framework, while analytically controlled, is broadly applicable to hidden-sector models with general vacuum, radiation, and matter contributions. We present the resulting predictions for PBH mass spectra, dark matter fractions, and gravitational-wave signals, highlighting parameter regions that remain open in current searches and motivating future probes.
\end{abstract}
\begin{document}
\preprint{MI-HET-866}
\title{Primordial Black Holes at the Junction}

\author{James B. Dent}
\email{jbdent@shsu.edu}

\affiliation{Department of Physics, Sam Houston State University, Huntsville, TX, USA}

\author{Bhaskar Dutta}
\email{dutta@tamu.edu}

\affiliation{Mitchell Institute for Fundamental Physics and Astronomy,\\Department of Physics and Astronomy, Texas A\&M
University, College Station, USA}

\author{Mudit Rai}
\email{muditrai@tamu.edu}

\affiliation{Mitchell Institute for Fundamental Physics and Astronomy,\\Department of Physics and Astronomy, Texas A\&M
University, College Station, USA}

\date{\today}

\maketitle

\newpage

\section{Introduction}
Primordial black holes (PBHs) -- black holes formed in the early universe not as the endpoint of standard stellar evolution -- are long-standing~\cite{Zeldovich:1967lct,Hawking:1971ei,Carr:1974nx,Hawking:1974rv} objects of interest as laboratories for general relativistic investigations as well as being candidates for dark matter (DM) (for reviews, see for example~\cite{Carr:2009jm,Carr:2021bzv,Green:2020jor,Carr:2024nlv}), as possible gravitational wave sources~\cite{Sasaki:2018dmp,Domenech:2021ztg,Yuan:2021qgz,LISACosmologyWorkingGroup:2023njw}, and as a viable candidate for massive seeds which could grow into the observed population of early supermassive black holes~\cite{Inayoshi:2019fun,Ziparo:2024nwh,Prole:2025snf}.

The formation of PBHs could arise through a variety of possible mechanisms including large primordial density fluctuations (possibly due to the underlying inflationary potential~\cite{Ozsoy:2023ryl}), from topolgical defects such as cosmic strings~\cite{Hawking:1987bn,Polnarev:1988dh,Vilenkin:2018zol} and domain walls~\cite{Deng:2016vzb,Ge:2019ihf,Liu:2019lul}, and from bubble collisions or collapse~\cite{Hawking:1982ga,Kodama:1982sf,Moss:1994iq,Khlopov:1998nm} associated with cosmological phase transitions. Recently there have been proposals for PBH formation during a cosmological phase transition, where particles are trapped in the false vacuum and form a non-topological soliton which eventually collapses to form a PBH~\cite{Hong:2020est,Kawana:2021tde,Baker:2021nyl,Kawana:2022lba,Lu:2024xnb,Dent:2025lwe}. Additionally,~\cite{Flores:2024lng} recently revisited PBH formation from false-vacuum bubbles whose gradient energy density in the bubble wall can lead to eventual collapse and PBH formation.

The criteria for the final collapse to a black hole can be of the form of a critical overdensity~\cite{Choptuik:1992jv,Evans:1994pj,Koike:1995jm,Niemeyer:1997mt,Kuhnel:2015vtw}, or a Schwarzschild criteria where the mass of the object under consideration is localized within its Schwarzschild radius.\footnote{A generalized form of the Schwarzschild criteria is the hoop conjecture~\cite{Klauder:1972je,Misner:1973prb,Ye:2025wif} which can accommodate non-spherical collapse.} Additionally, preceding the collapse into a black hole, a false-vacuum bubble can have a `turning point' signified, for example, by the location where the system's energy is equivalent to its potential~\cite{Flores:2024lng}, or where a balance of pressures between the true and false vacua, determined by a minimization condition on the Helmholtz free energy, can stabilized the system before additional contributions cause the collapse to a black hole~\cite{Hong:2020est,Kawana:2021tde,Marfatia:2021hcp,Lu:2024xnb,Dent:2025lwe}.

We emphasize that PBH formation through a first-order phase transition is particularly appealing, as it is naturally accompanied by a gravitational-wave (GW) signal generated during the transition~\cite{Kosowsky:1991ua,Kosowsky:1992rz,Apreda:2001us,Grojean:2006bp,Huber:2008hg,Cai:2017cbj,Weir:2017wfa,Caprini:2009yp,Caprini:2015zlo,Caprini:2018mtu,Caprini:2019egz,Caprini:2024hue,Mazumdar:2018dfl,Caldwell:2022qsj,Hindmarsh:2020hop}, the strength of which is directly controlled by the available vacuum energy. 

In this work we compare and generalize these different collapse criteria for first order phase transitions, where false-vacuum patches can survive and generate inhomogeneities which can lead to conditions favorable for PBH formation. Specifically, we study the Israel junction conditions~\cite{Israel:1966rt} as they apply to nucleated false vacua. The junction conditions, which follow from the Einstein equations (see Appendix \ref{sec:Junction_conditin}), specify how two spacetimes can be consistently matched across a thin hypersurface. While they have long been studied in the context of phase transitions~\cite{Berezin:1982ur, Blau:1986cw}, their application to PBH formation in a vacuum-dominated cosmology is only recent~\cite{Flores:2024lng}.


Here, we take a more general approach by considering a hidden sector described by a scalar field obeying a polynomial potential with general energy density contributions (radiation, matter, and vacuum) undergoing a phase transition in a radiation-dominated universe. We assume that the hidden sector is in thermal equilibrium at temperature $T_h$, while the Standard Model (SM) bath has temperature $T_{\text{SM}} = T_h/\xi$ with $\xi \leq 1$. We consider the case where the SM contribution to the Hubble rate dominates the universe's expansion.

As a first concrete scenario, we apply the junction conditions to investigate the collapse of exotic objects which display a particle trapping mechanism, such as Fermi-balls (FBs), into PBHs.\footnote{Although there exist other scenarios where PBHs are formed via particle trapping~\cite{Baker:2021nyl,Dent:2025lwe}, we will take the Fermi-ball case as the exemplar of this formation category.} Remarkably, since the junction conditions depend only on the energy difference between the false and true vacuum, PBH formation can proceed even when the majority of this energy is sequestered in Fermi-balls rather than in the vacuum itself. This leads to a distinctive trade-off: the corresponding gravitational-wave GW signal is shifted towards higher frequency compared to the FB case relative to the vacuum only case. Importantly, our approach is entirely novel with respect to Fermi-balls, as it bypasses the conventional pressure-balancing arguments commonly employed in the literature~\cite{Hong:2020est,Kawana:2021tde,Marfatia:2021hcp,Lu:2024xnb,Dent:2025lwe}, providing a more general and robust description of Fermi-ball–mediated PBH formation.

In our second example, we emphasize that the junction conditions are generic and as an illustrative application, we consider the case of multiple phase transitions~\cite{Dent:2024bhi}, showing that such a setup yields correlated GW signatures as well as correlated PBH abundances across the different transitions.

Our choice of models allows for analytic control, but the framework is quite general and can be extended to a wide class of scenarios. In particular, while we focus on the case where the hidden sector provides only a subdominant contribution to the energy density, the setup can be readily generalized to situations in which the hidden sector plays a significant role in driving the Hubble expansion. 


Using these two scenarios, We will calculate the GW strengths  along with the possible PBH mass spectrum and current fractional dark matter contribution characterized by the parameter $f_{\rm PBH} \equiv \rho_{\rm PBH}/\rho_{\rm DM}$, where the ratio is of the current PBH and dark matter densities. 


The remainder of this work is as follows. In Sec.~\ref{sec:PBH_FB} we provide details of the Israel junction conditions appraoch to PBH formation and apply it to a general false vacuum as well as to the case of Fermi-balls. Sec.~\ref{sec:PBH_multi} discusses the formation of PBHs within a multiple phase transition scenario. The survival probability is calculated in Sec.~\ref{sec:survival}, which allows us to calculate the PBH dark matter fraction, $f_{\rm PBH}$. In Sec.~\ref{sec:results} we show the results for the GW spectra and their correlated results in the $f_{\rm PBH}-M_{\rm PBH}$ plane. Finally, we summarize our work in Sec.~\ref{sec:conclusions}. In Appendix~\ref{app:hidden_sector_model} we provide details of the hidden sector model undergoing a phase transition and describe the Israel junction conditions. In Appendix~\ref{sec:App_FB} we review the standard analysis of the Fermi-ball scenario using the free energy, and then discuss the trapping scenario from the turnaround approach in Appendix~\ref{app:trapping}. In Appendix~\ref{app:overdensity} we include a brief discussion of the overdensity criteria for PBH formation. In Appendix~\ref{app:GW_param}, we briefly describe the gravitational wave parametrization used in this work.
\section{PBH Formation : Fermi-balls \label{sec:PBH_FB}}

The Israel junction conditions provide a general framework for analyzing the dynamics of thin hypersurfaces separating different spacetime regions. These conditions are not limited to vacuum bubble walls, but can also be applied to scenarios where additional matter degrees of freedom are trapped inside the false vacuum. In particular, such configurations may give rise to non-topological solitonic objects such as Fermi-balls~\cite{Hong:2020est} or Q-balls~\cite{Coleman:1985ki, Frieman:1988ut}.

Fermi-balls are bound states that arise when fermions become localized inside regions of false vacuum, where their mass is suppressed relative to the true vacuum. They are trapped in the false vacuum if their mass gain in traversing to the true vacuum is too large for them to access due to their thermal energy being too small. As the phase transition proceeds, the fermions accumulate in these false-vacuum domains, and the resulting Pauli pressure can counterbalance the vacuum energy difference and surface tension, stabilizing the configuration (see Appendix~\ref{sec:App_FB} for more details on this point). These objects are examples of non-topological solitons whose stability is ensured by a conserved fermion number, rather than by topological arguments.

In the setup we consider, a fermionic field $\chi$ couples to the scalar field $\phi$ responsible for the phase transition, with interaction strength $g_\chi$. Inside the false vacuum, $\chi$ becomes trapped and accumulates as the transition progresses, leading to the formation of Fermi-balls. 
The Lagrangian relevant to our considerations is given as,
\[
\mathcal{L} \subset \Bar{\chi}(i\,\slashed \partial -m)\chi -g_\chi\,\phi\,\bar{\chi}\chi - V(\phi,T)
\]
where the potential is given by, 
\[
V(\phi,T) \approx D(T^2-T_0^2)\phi^2 - E\,T \phi^3 + \frac{\lambda}{4} \phi^4,
\label{eq:V}
\]
Successful trapping requires that the mass difference between the true vacuum and the false vacuum be larger than the critical temperature i.e.,
\[
g_\chi \phi_c \gg T_c \implies g_\chi \gg \frac{\lambda }{2E}
\]
 This condition imposes a stringent constraint on the coupling strength. Furthermore, the bare mass term must remain small compared to other dimensionful parameters, i.e. $m \ll T_0$, where $T_0 $ is the scale of the hidden sector and $(mT)^2 \ll \rho_V$, in order for particles to remain trapped.\footnote{Otherwise, large bare masses would render the masses in the true and false vacua nearly identical, preventing efficient trapping.}  

In addition to Fermi-ball models, we note that there exist scenarios in the literature where particles are trapped within the false vacuum~\cite{Baker:2021nyl} without forming Fermi-balls. In Appendix~\ref{app:trapping}, we apply the Israel junction conditions to such cases and demonstrate that PBH formation can still be consistently analyzed. As expected, for a given amount of energy stored in the false vacuum, the resulting PBH abundance and gravitational-wave signal are comparable to those obtained in the Fermi-ball scenario.

The effective Yukawa interaction length is given by~\cite{Marfatia:2021hcp},  
\begin{equation}
L_\phi = \left.\left(\frac{d^2V}{d\phi^2}\right)^{-1/2}\right|_{\phi=0} 
= \frac{1}{\sqrt{2D\,(T^2-T_0^2)}} 
= \frac{\sqrt{\lambda}}{E\,T\,\sqrt{\kappa(T)}} \, .
\end{equation}

The energy density trapped inside the false vacuum can be obtained from the free energy,  
\begin{equation}
\mathcal{E} = F - T\,\frac{\partial F}{\partial T} \, , 
\qquad 
\rho_{\rm in} = \frac{\mathcal{E}}{V_{\rm in}}, 
\qquad V_{\rm in} = \frac{4\pi}{3}\,R^3 \, .
\end{equation}
The free energy is given by, 
\begin{equation}
F = \left(f_Q(R,T) + \rho_V - \frac{\pi^2}{90}\,T^4\right)\frac{4\pi R^3}{3} \, ,
\end{equation}
where,\footnote{Including Yukawa and temperature-dependent effects and we neglect subleading terms of order $\mathcal{O}(m/T)$ and $\mathcal{O}(mT/\sqrt{\rho_V})$.}
\begin{equation}
f_Q(r,T) = \left(\frac{3}{2\pi}\right)^{2/3}\,\frac{9\,Q_{FB}^{4/3}}{16\,r^4}
\left(1+\frac{4\pi}{9}\left(\frac{2\pi}{3}\right)^{1/3}\frac{r^2T^2}{Q_{FB}^{2/3}}\right)
- \frac{1}{2}\left(\frac{3g_\chi}{4\pi}\,\frac{Q_{FB}\,L_\phi}{r^3}\right)^2 \, .
\end{equation}
which leads to  
\begin{equation}
\rho_{\rm in} = \rho_{\rm rad,in} + \rho_V + \rho_{\rm FB} \, ,
\end{equation}
with the individual contributions
\begin{align}
\rho_V & =  \frac{\lambda\,v^4}{4},\\ 
\rho_{\rm rad,in} &= \frac{\pi^2}{30}\,g_{\rm eff}\,T^4 \,, \\
 \rho_{\rm FB} &= f_Q(R,T) - T\,\frac{\partial f_Q(R,T)}{\partial T} \, .
\end{align}

We will apply the Israel junction condition (cf. Appendix \ref{sec:Junction_conditin}) across the interior and exterior of the false vacuum bubbles. We find that the radiation contribution cancels out when computing the energy density difference $\Delta \rho$ between the false and true vacua. In addition to the vacuum energy $\rho_V$, the trapped contribution from Fermi-balls appears through $f_Q(R,T)$.
This leads to the Fermi-ball energy density
\begin{equation}
\rho_{FB}(r,T) = \left(\frac{3}{2\pi}\right)^{2/3}\,\frac{9\,Q_{FB}^{4/3}}{16\,r^4}
\left(1-\frac{4\pi}{9}\left(\frac{2\pi}{3}\right)^{1/3}\frac{r^2T^2}{Q_{FB}^{2/3}}\right)
- \frac{1}{2}\left(\frac{3g_\chi}{4\pi}\,\frac{Q_{FB}\,L_\phi}{r^3}\right)^2 \, ,
\end{equation}
where $Q_{FB}$ denotes the conserved charge associated with the Fermi-ball.  
 We can condense the notation by defining the functions of $r$,
\begin{align}
    & f_Q(r) = \frac{a}{r^4} + \frac{d}{r^2}-\frac{b}{r^6} \\ 
    & \rho_{FB}(r) = \frac{a}{r^4}  -\frac{d}{r^2}-\frac{b}{r^6},
\end{align}
where $a = \left(\frac{3}{2\pi}\right)^{2/3}\,\frac{9\,Q_{FB}^{4/3}}{16} $, $b = \frac{1}{2}\,\left(\frac{3\,g_\chi}{4\pi}\,Q_{FB}\,L_\phi\right)^2$, $c =\rho_V$, and $d = \left(\frac{3}{2\pi}\right)^{1/3} \frac{\pi\,T^2}{4}Q_{FB}^{2/3}$.\footnote{Similarly, we could write down an expression for Q-balls formed due to scalar field condensation as well.}

The junction condition becomes~\cite{Berezin:1982ur},
\[
M = \frac{4\,\pi}{3}\Delta\,\rho\,r^3 - \pi\frac{\sigma^2}{M_{pl}^2}
\,r^3+4\pi\sigma\,r^2\sqrt{1-\frac{\rho_{in}}{3\,M_{pl}^2}r^2+\dot{r}^2},
\label{eq:junction1}
\]
where $\Delta\,\rho = \rho_{in} - \rho_{out} = \rho_V + \rho_{FB}$, and the dot notation is $\dot{} \equiv d/d\tau$, with $\tau$ being the physical time as measured along the wall trajectory.
The above equation can be re-expressed in terms of dimensionless variables by defining,
\begin{align}
    z^3 = \left(\frac{4\pi\rho_V}{3M\cos^2\theta}\right)r^3,\qquad\tau' =  \left(\frac{\sqrt{\rho_V/3}}{\sin(2\theta) M_{pl}}\right)\tau,\qquad\overline{M} = \frac{4\pi\,M_{pl}^3}{\sqrt{\rho_V/3}},
    \label{eq:dimless_vars}
\end{align}
where $\overline{M}$ is the maximum mass allowed for the collapsing PBH, and $\theta$ is defined through the expression
\begin{align}
    \theta = \tan^{-1}\left(\frac{\sigma}{2\,M_{pl}\sqrt{\rho_V/3}}\right)
    \label{eq:theta}
\end{align}
We then find that Eq.~(\ref{eq:junction1}) can be written as
\begin{align}
\left(\frac{d\,z}{d\tau'}\right)^2 + U(z) = E,
\end{align}
where $U(z)$ is given by, 
\begin{align}
U(z) =& -\left(\frac{1-z^3}{z^2}\right)^2 
- \frac{4\sin^2\theta}{z} 
-  4\sin^2\theta\,\cos^2\theta \, \frac{\rho_{rad}}{\rho_V}\, z^2 \\ \nonumber
& - z^2 \cos^2\theta\,\frac{\rho_{FB}(z)}{\rho_V}
\left(-\frac{2}{z^3} + \cos^2\theta \, \frac{\rho_{FB}(z)}{\rho_V}\right) 
- 2 z^2 \cos^2\theta \, \frac{\hat{\rho}_{FB}(z)}{\rho_V} \, .
\end{align}
along with the expressions
\begin{align}
\cos^2\theta\,\frac{\rho_{FB}(z)}{\rho_V} &=  \frac{3}{4\,z^4}\left(\frac{Q_{FB}}{M}\right)^{4/3}\left(\frac{3\,\pi^2\,\rho_V}{\cos^2\theta}\right)^{1/3} -\frac{T^2}{2\,z^2}\left(\frac{\pi^2\,Q_{FB}\,\cos\theta}{M}\right)^{2/3}\left(\frac{1}{3\,\rho_V}\right)^{1/3} \\ \nonumber & - \left(\frac{g_\chi\,L_\phi\,Q_{FB}}{\sqrt{2}\,M\,\cos\theta}\right)^2\,\frac{\rho_V}{z^6}
\\
\cos^2\theta\,\frac{\hat{\rho}_{FB}(z)}{\rho_V} &= \frac{1}{4\,z^4}\left(\frac{3\,Q_{FB}}{M}\right)^{4/3}\left(\frac{\pi^2\,\rho_V}{\cos^2\theta}\right)^{1/3} - \left(\frac{g_\chi\,L_\phi\,Q_{FB}}{\sqrt{2}\,M\,\cos\theta}\right)^2\,\frac{\rho_V}{z^6}
\\
E &= -4\sin^2\theta\,\left(\frac{\cos\theta\,\overline{M}}{M}\right)^{2/3} - T^2\,\left(\frac{\pi^2\,Q_{FB}}{M}\right)^{2/3}\left(\frac{1-\frac{\gamma^2}{4}}{3\,\rho_V}\right)^{1/3},
\end{align}


In terms of the usual Fermi-ball equation parameters, we have
\begin{align}
\frac{\rho_{FB}(z)}{\rho_V} &= \left(\frac{3\,a_1}{z^4}-\frac{3\,a_2}{z^2} - \frac{a_0}{z^6}\right)
\\
\frac{\hat{\rho}_{FB}(z)}{\rho_V} &= \left(\frac{3\,a_1}{z^4}- \frac{a_0}{z^6}\right)
\\
E &= -4\sin^2\theta\,\cos^{2/3}\theta\left(\frac{\overline{M}}{M}\right)^{2/3} - 6\,a_2\cos^2\theta
\end{align}
where the various $a_i$ are given by
\begin{align}
    & a_1 =\frac{a}{3\, \Xi ^4 c} = \frac{3^{1/3} \, \pi^{2/3} \, Q_{FB}^{4/3} \, \rho_V^{1/3}}{4 M^{4/3} \cos^{8/3}\theta},\quad a_0 = \frac{b}{\Xi ^6 c} = \frac{g_\chi^2 L_\phi^2 Q_{FB}^2 \rho_V }{2\,\cos^4\theta\, M^2},\quad \\ & a_2= \frac{d}{3\, \Xi ^2 c} = \frac{T^2}{6\,(3\,\rho_V)^{1/3}}\left(\frac{\pi^2\,Q_{FB}\,\cos\theta}{M}\right)^{2/3},
\end{align}
with $\Xi = \left(\frac{3 M \cos ^2\theta }{4 \pi  \rho_V}\right)^{1/3}$. The turning point, defined by $U(z) =E$ characterizes the point at which the radial velocity vanishes. Physically, this represents
the moment when the shell comes to a halt before collapsing inward and hence determines whether the false vacuum patch
collapses into a black hole. The equation reduces to, 
\[
\left(1-\frac{1}{z^3}+\cos^2\theta\frac{\rho_{FB}(z)}{\rho_V}\right)^2 + \frac{4\sin^2\theta}{z^2}\left(\,z^2\,\left(\frac{1}{z^3} + \cos^2\theta\frac{\rho_{rad}}{\rho_V}\right) - \cos^{2/3}\theta \left(\frac{\overline{M}}{M}\right)^{2/3}\right) = 0
\label{eq:z_TP_fermi}
\]

We point out that the above equations exhibit a hierarchy in terms of small 
$\theta$ or, equivalently, small 
$\tfrac{\sigma}{\sqrt{\rho_V}\,M_{pl}}$. 
In Eq.~(\ref{eq:junction1}), the terms 
$\{M,\,\tfrac{4\pi r^3}{3}\Delta\rho\}$ and the dynamical term 
$4\pi\sigma r^2\dot{r}$ are all of order $O(1)$ in $\theta$ once written 
in dimensionless form, whereas the remaining contributions are suppressed 
to $O(\theta^2)$. Retaining only the leading-order terms yields the simplified 
dynamical equation
\begin{equation}
\left(\frac{dz}{d\tau'}\right)^2 = 
z^2\left(1-\frac{1}{z^3}+\frac{\rho_{FB}(z)}{\rho_V}\right)^2 
+ O(\theta^2)\,.
\label{eq:z_TP_trunc}
\end{equation}
The turning point, defined by $dz/d\tau' = 0$, then satisfies
\begin{equation}
1-\frac{1}{z^3}+\frac{\rho_{FB}(z)}{\rho_V}=0\,.
\end{equation}
This agrees with Eq.~(\ref{eq:z_TP_fermi}) when restricted to $O(1)$ 
and neglecting the $\sin^2\theta$ term. Another notable point to mention is that in the absence of Fermi-ball energy, $\rho_{FB} =0$, we recover the turning point equation for the vacuum energy scenario given in Eq.~(\ref{eq:zTP_vac}) where $z_{TP} = 1 + O(\theta^2)$, which matches the scenario discussed in~\cite{Flores:2024lng}.

We emphasize that this derivation 
does not invoke force-balance arguments, which are often used in earlier 
studies of Fermi-balls and lead instead to an alternative
condition (cf. Appendix~\ref{sec:App_FB}).
Another scenario considered in the literature involves fermions being trapped inside the false vacuum patch~\cite{Baker:2021nyl} due to kinematics. 
It has been shown that such trapped fermions can lead to PBH formation by satisfying the Schwarzschild condition. 
In Appendix~\ref{app:trapping}, we demonstrate that the PBH collapse can be obtained by applying the Israel--Junction conditions, 
allowing us to determine both PBH formation and the associated gravitational wave signal.

\section{PBH formation :  Mulitple transitions\label{sec:PBH_multi}}
In this section, we discuss the formation of PBHs during a sequence of multiple phase transitions~\cite{Dent:2024bhi,Barni:2024lkj,Buen-Abad:2023hex}. In the multiple phase transition scenario under consideration, following the completion of the first phase transition, energy injection in the early universe~\cite{Dutta:2009uf, Lyth:1998xn, Kibble:1976sj} triggers reheating in the hidden sector, dynamically leading the hidden-sector field to undergo subsequent transitions. Among these, the first and third transitions follow the conventional symmetry-breaking pattern, while the second is an inverted transition, proceeding from a broken to an unbroken phase. In this work, we analyze the PBH formation associated with the first and third transitions. The criteria for PBH formation in this scenario can only be derived from the turning-point equations. Unlike the Fermi-ball case, no matter contribution arises here. A detailed investigation of the PBH formation in an inverted transition is more involved and is deferred to future studies.

As discussed in \cite{Dent:2024bhi}, Phase 1 and Phase 3 are structurally similar, differing primarily in the Hubble parameter at which the transition occurs. It is crucial to ensure that energy injection takes place only after the false vacuum region has collapsed into a black hole; otherwise, the injection may convert the false vacuum into the true vacuum, thereby halting the collapse and preventing black hole formation. Thus we need,
\[
t_\delta -t_N > t_{collapse}
\]
where $t_\delta$ is the time of energy injection and $t_N$ is the nucleation time.
The junction conditions for both the transitions would be the same with the corresponding Hubble parameter, 
\[
M = \frac{4\,\pi}{3}\Delta\,\rho\,r^3 - \pi\frac{\sigma^2}{M_{pl}^2}
\,r^3+4\pi\sigma\,r^2\sqrt{1-\frac{\rho_{in,i}}{3\,M_{pl}^2}r^2+\dot{r}^2}
\label{eq:junction2}
\]
As in the previous section, using the definitions
\begin{align}
z^3 = \left(\frac{4\pi\rho_V}{3M\cos^2\theta}\right)r^3\;\;\;;\;\;\;  \overline{M} = \frac{4\pi M_{pl}^3}{\sqrt{\rho_V/3}} = \frac{8\pi M_{pl}^3}{v^2}\sqrt{\frac{3}{\lambda}}  
\end{align}
where $\overline{M}$ is the maximum mass allowed for the collapsing PBH, 
Eq.~(\ref{eq:junction2}) can be re-expressed in terms of dimensionless variables, 
\[
\left(\frac{d\,z}{d\tau'}\right)^2 + U_i(z) = E,
\]
where $U(z)$ is given by, 
\[
U_i(z) = -\left(\frac{(1-z^3)}{z^2}\right)^2 - \frac{4\,\sin^2\theta}{z}-4\,\sin^2\theta\,\cos^2\theta\,\left(\frac{\rho_{in,i}}{\rho_V}-1\right)\,z^2.
\]
where $\theta$ is defined in Eq.(\ref{eq:theta}). The index $i =\{1,3\}$ denotes which phase transition is being considered and 
\[
E = -4\,\sin^2\theta\,\left(\frac{\cos\theta\,\overline{M}}{M}\right)^{2/3},
\]
This leads to the equations for the turning points satisfying $U_i(z)=E$ given by,
\[
\left(1-\frac{1}{z^3}\right)^2 + \frac{4\sin^2\theta}{z^2}\left(\,z^2\,\left(\frac{1}{z^3} + \cos^2\theta\,\frac{\rho_{rad,i}}{\rho_V}\right) - \cos^{2/3}\theta \left(\frac{\overline{M}}{M}\right)^{2/3}\right) = 0
\label{eq:zTP_multi}
\]
The solution for the turning point value $z = z_{TP,i}$ for the $i$'th transition is given by $z_{TP,i} \sim 1$ for both transitions as expected, since the primary difference between them arises only through the Hubble parameter, which affects the result at subleading order. The distinction becomes manifest phenomenologically: because the transitions occur at different values of the Hubble scale, they generate distinct GW spectra as well as different values of $f_{\rm PBH}$.

\section{Survival Probability}
\label{sec:survival}

We aim to compute the survival probability of false-vacuum patches at the turning point, $t_{TP}$. 
To this end, we determine the characteristic size of a false-vacuum patch, $r_{TP}$, obtained from the junction conditions. 
We then relate this quantity to the late-time relative relic abundance of the PBHs that survive until today, defined as $f_{\mathrm{PBH}}=\rho_{\mathrm{PBH}}/\rho_{\mathrm{DM}}$.

The survival probability of a false vacuum patch is given by \cite{Athron:2023xlk},
\[
f_{FV}(t) = e^{-I(t)},
\]
with the exponent $I(t)$  given by the integral form
\[
I(t) = \int^t_{t_c} dt' \Gamma(t') \left(\frac{a(t')}{a(t)}\right)^3\,\frac{4\pi}{3}R^3(t,t')
\label{eq:I(t)}
\]
where $R(t,t')$ is the radius at time $t$ of a true vacuum bubble nucleated at time $t'$ given by,
\[
R(t,t') = v_w\,\int^t_{t'} dt''  \frac{a(t'')}{a(t)},
\]
and $\Gamma(t)$ is the rate of nucleation of the true vacuum. 
The mean true bubble separation is given by \cite{Athron:2023xlk}, 
\[
r_{sep}(t) = n_{tv}(t)^{-1/3}
\label{eq:rsep}
\]
where the density is provided by the form
\[
 n_{tv}(t) = \int_{t_c}^t \Gamma(t') f_{FV}(t') \left(\frac{a(t')}{a(t)}\right)^3\,dt'
 \label{eq:ntv}.
\]
We will relate the bubble separation at the turning point, $t=t_{TP}$, to the radius of the false patch, $r_{sep}(t_{TP}) \approx 2\,r_{TP}$.
The fractional relic abundance for PBHs is given by \cite{Marfatia:2021hcp}, 
\[
f_{\rm PBH} = \frac{n_{FV,0}\,M_{\rm BH}}{\rho_{\rm DM}}
\]
where 
\[
n_{FV,0} = n_{FV}(t_{TP}) \left(\frac{s_0}{s(T_{TP})}\right)
\]
by using entropy conservation. We estimate the false vacuum density by following the reverse time description developed in \cite{Lu:2022paj}, 
\[
n_{FV}(t) = \int_{t}^{t_e} \Gamma_r(t') (1-f_{FV}(t')) \left(\frac{a(t')}{a(t)}\right)^3\,dt'
\label{eq:nfv}
\]
\subsection{Exponential nucleation}
The estimation of the false-vacuum bubble probability can be significantly simplified in our model, as the nucleation rate can be approximated by an exponential of the form
\[
\Gamma(t) \approx \Gamma_*\,e^{\beta(t-t_*)} \, ,
\]
where $t_* \in (t_c, t_e)$, where $t_c$ corresponds to the start of the transition (corresponding to critical temperature),  $t_e$ is the time corresponding to the end of the transition and $\beta = -\left. (dS/dt)\right|_{t_*}$ characterizes the rate of the phase transition. In our scenario, we take $t_* \approx t_N$, the nucleation time. Within this simplified framework, the integral\footnote{The following expression is derived neglecting the scale factor. In a radiation-dominated cosmology, including the scale factor introduces only a negligible correction in the limit $\beta/H \gg 1$.}  $I(t)$ can be approximated as
\[
I(t) \approx \frac{8 \pi v_w^3 \Gamma_*}{\beta^4}\, {\rm exp}\left({\frac{\beta}{2 H_*} \left(\frac{t}{t_*}-1\right)}\right) \equiv I_*\, {\rm{exp}}\left({\frac{\beta}{2 H_*} \left(\frac{t}{t_*}-1\right)}\right),
\]
where $I_* \approx 1.238$ \textcolor{red}~\cite{Lu:2022paj}. Substituting this into Eq.~(\ref{eq:ntv}) gives
\[
n_{TV}(t_{TP}) = \int_{t_c}^{t_{TP}} dt'\, \Gamma_* e^{\beta (t'-t_*)} \, e^{-I_* e^{\beta(t'-t_*)}} \left(\frac{t'}{t_{TP}}\right)^{3/2}.
\]
Using the change of variable $y = e^{\beta (t'-t_*)}$, we obtain
\[
n_{TV}(t) = \left(\frac{t_*}{t}\right)^{3/2} \frac{\Gamma_*/H_*}{\beta/H_*} \int_{y_c}^{y} dy'\, e^{-I_* y'} \left(1 + \frac{2 \ln y'}{\beta/H_*}\right)^{3/2}.
\]
Now we can plug in the turning point radius since we estimate it to be roughly the separation distance between true-vacuum regions, $
2 r_{TP} = n_{TV}(t_{TP})^{-1/3}$.
This leads to
\[
\frac{t_{TP}}{t_*} = \left(\frac{\beta/H_*}{(8\pi)^{1/3} v_w} \, 2 r_{TP} H_*\right)^2 I_*^{2/3} \left( \int_{y_c}^{y_{TP}} dy\, e^{-I_* y} \left(1 + \frac{2 \ln y}{\beta/H_*}\right)^{3/2} \right)^{2/3}.
\]
The integral in parentheses is fairly insensitive to $t_{TP}$, so we may approximate
\begin{equation}
t_{TP} \approx 1.15\, t_* \left(\frac{\beta/H_*}{v_w} r_{TP} H_* \right)^2 \, .
\label{eq:t_TP}
\end{equation}

For an exponential nucleation rate, the false-vacuum number density in Eq.~(\ref{eq:nfv}) simplifies significantly, since $\Gamma_r$ is given by~\cite{Lu:2022paj}, 
\[
\Gamma_r(t) \approx 
\frac{\bigl(\beta_*\, I_*\, e^{\beta_*(t-t_*)}\bigr)^4}{192\, v_w^3}\,
e^{-I_* e^{\beta_*(t-t_*)}}
\]
In the reverse-time description \cite{Lu:2022paj}, one can approximate that a false-vacuum patch nucleated at time $t'$ reaches $r_{TP}$ at $t_{TP}$. For fast transitions ($\beta/H > \mathcal{O}(100)$), the scale factor can be safely neglected, so that $r/v_w \approx t_{TP}-t'$\footnote{Although in our scenario the false patch collapses to $r_{\rm Schwz}$, since $r_{\rm Schwz} \ll r_{TP}$, it can be effectively set to zero.}.  
Following \cite{Marfatia:2024cac}, this yields
\[
n_{FV}(t_{TP}) \approx H_*^3 \frac{(\beta/H_*)^3}{192\, v_w^3} x^4 e^{-x} (1 - e^{-x}),
\]
where
\[
x = e^{\frac{\beta_* r_{TP}}{v_w}}.
\]

Combining the above results, the fractional relic abundance of PBHs can be expressed as\footnote{We estimate the wall velocity semi-analytically following \cite{Ellis:2020awk}. This method is well-suited for polynomial-like potentials and yields errors of less than an order of magnitude for $f_{PBH}$.}
\begin{align}
f_{\rm PBH} &\approx \frac{M_{\rm PBH}}{\rho_{\rm DM}} \frac{s_0}{s_{TP}}\, n_{FV}(t_{TP}) \nonumber \\
&\approx 14.4 \times \left(\frac{M_{\rm PBH}}{10^{18}\,\rm g}\right)
\left(\frac{\beta/H}{10^3}\, \frac{1}{v_w}\right)^3
\left(\frac{\xi\,T_*}{20\,\rm MeV}\right)^3\,\left(\frac{T_*\,\sqrt{g_{*,\rho,SM}(T_*)+g_h\,\xi^4}}{T_{TP}\,(g_{*,S,SM}(T_{TP})+g_h\,\xi^3)}\right)^3
x^4 e^{-x} (1 - e^{-x}) \nonumber \\ 
&\approx 0.2 \times \left(\frac{M_{\rm PBH}}{10^{18}\,\rm g}\right)
\left(\frac{\beta/H}{10^3}\, \frac{1}{v_w}\right)^3
\left(\frac{T_*}{20\,\rm MeV}\right)^3\,
x^4 e^{-x} (1 - e^{-x})
\end{align}
where in the last line we set $\xi =1$ and assumed $T_{TP}\sim T_*\sim 20\,\rm{MeV}$ to get a sense of scales, and we have the combination
\[
x^4 e^{-x} (1-e^{-x}) \sim
\begin{cases}
0.23, & \text{for } \frac{\beta}{H_*}\,\frac{r_{TP} H_*}{v_w} \ll 10^{-2}, \\[2mm]
\mathcal{O}(1), & \text{for } \frac{\beta}{H_*}\,\frac{r_{TP} H_*}{v_w} \sim \mathcal{O}(1),
\end{cases}
\label{eq:TPdep_fpbh}
\]
and
\[
x = \exp\!\Bigg(\frac{\beta/H_*}{v_w} r_{TP}\, H_*\Bigg).
\]
For fast transitions, $t_{TP} \sim \mathcal{O}(1)\,t_*$, so from Eq.~(\ref{eq:t_TP}) we see that we generally lie in the regime where the exponential factor is $\frac{\beta}{H_*}\,\frac{r_{TP}\,H_*}{v_w} \sim \mathcal{O}(1)$.

\section{Results}
\label{sec:results}
In this section, we explore the phenomenological implications of our analysis by highlighting the interplay between the gravitational wave signature of the phase transitions and the resulting PBH abundance. The GW signal is described using a semi-analytical parametrization that accounts for contributions from bubble wall collisions, sound waves, and turbulence, with further details given in Appendix~\ref{app:GW_param}. We find that these observables are strongly correlated -- the strength and dynamics of the phase transitions not only govern the amplitude and shape of the GW signal but also directly influence the formation and late-time abundance of PBHs. This complementarity provides a powerful window into probing the underlying physics producing the transitions.

\begin{figure}[h!]
\centering

\includegraphics[width=\linewidth]{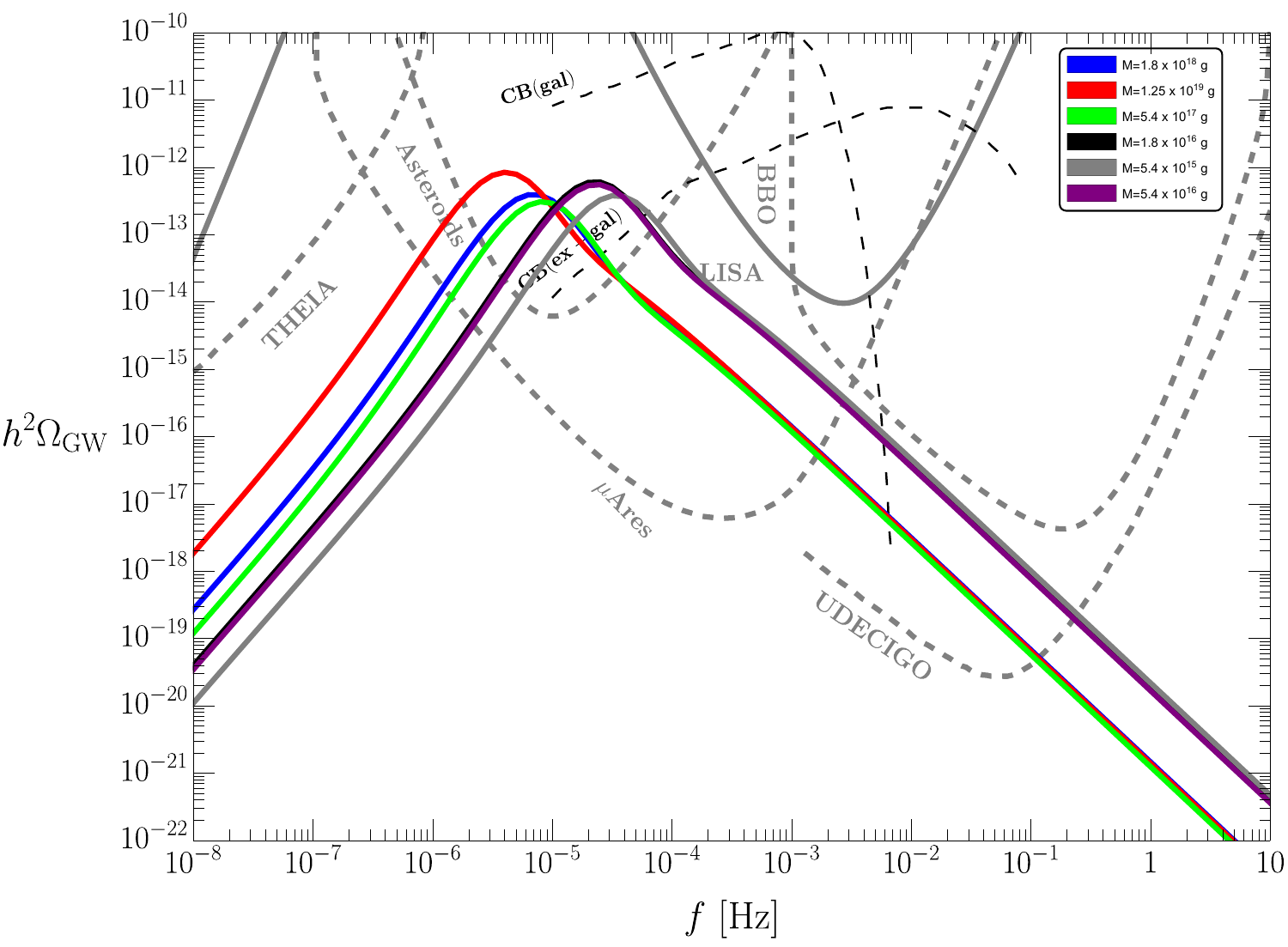} 

 \caption{\justifying GW spectrum, $h^2\Omega_{\rm GW}$ vs $f\left[\rm Hz\right]$ for different masses and scales (red, blue and green curves corresponds to $T_0 = 20 \,\rm{MeV}$ and black, gray and violet curves corresponds to $T_0 = 100 \,\rm{MeV}$). All of the six scenarios corresponds to $\rho_V/\rho_{rad} \approx 0.05$. Also
shown are the PLIS curves for upcoming experiments LISA (solid gray) and proposed experiments µAres, asteroid laser ranging, and
BBO (dashed gray). The PLIS curves for LISA and BBO are adopted from \cite{Schmitz:2020syl, Batell:2023wdb} but
scaled to observation times of 3 yrs for LISA \cite{Caprini:2019egz} and 4 yrs for BBO \cite{Crowder:2005nr}.
The µAres PLIS is taken from \cite{Sesana:2019vho}, scaled to SNR = 1. For the asteroid ranging proposal, we
adopt the strain sensitivity given in \cite{Fedderke:2021kuy} and calculate the PLIS curve using the procedure outlined
in \cite{Caprini:2019egz} for SNR = 1 with an assumed experiment duration of 7 yrs. For UltimateDECIGO (UDECIGO) we have adopted the PLIS in \cite{Braglia:2021fxn}. Black dashed lines represent foregrounds
from galactic and extragalactic compact binaries (CB)\cite{Robson:2018ifk, Farmer:2003pa}. }
\label{fig:GW_multiple_masses}
\end{figure}
\begin{figure}[h!]
\centering
\includegraphics[width=\linewidth]{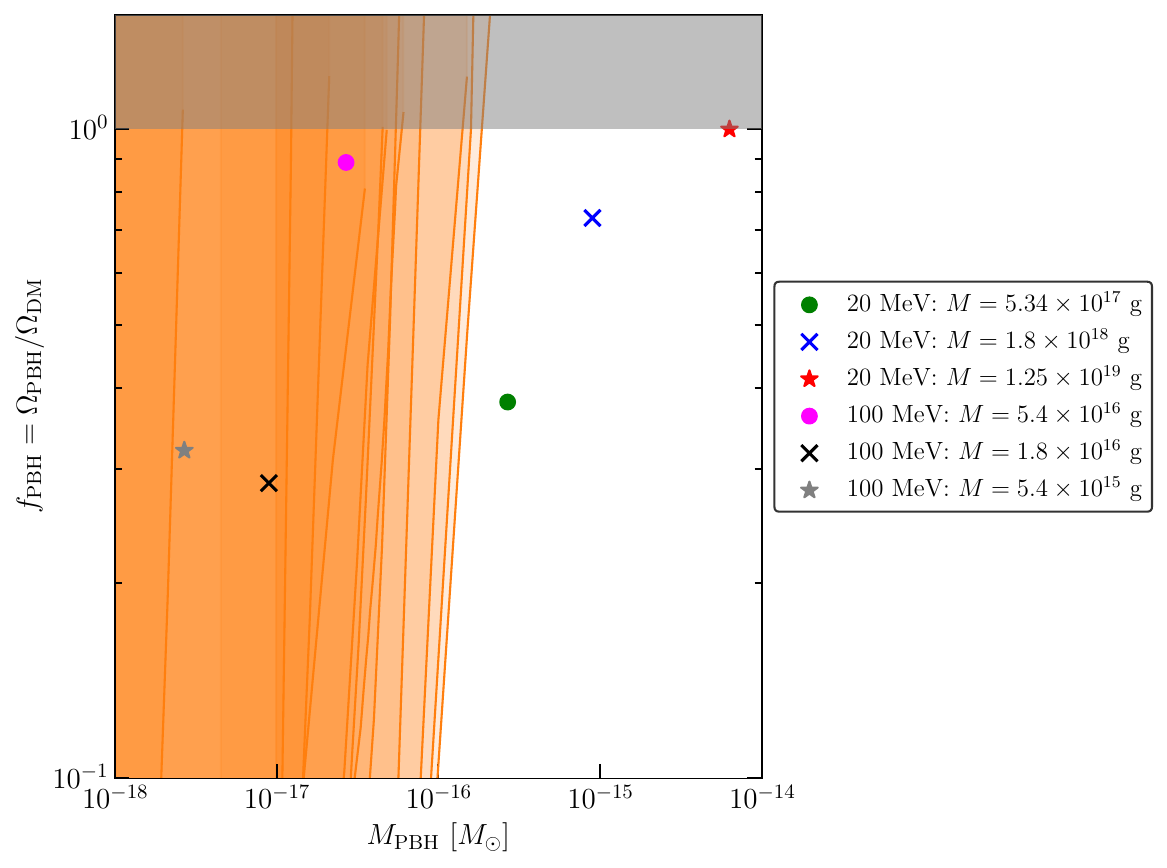} 
 \caption{\justifying$f_{\rm PBH}$ as a function of $M_{\rm PBH}$ for two different benchmark hidden sector temperature scales and a range of $M_{\rm PBH}$ for each temperature. The mass value is a parameter choice which can be determined by choosing the corresponding value of the other model parameters. This scenario corresponds to hidden sector where the PBH's are sourced from the vacuum energy difference between the false and true vacua. 
The orange shaded region indicates the exclusion from PBH evaporation constraints~\cite{Green:2020jor}.}
\label{fig:PBH_multi}
\end{figure}


\subsection{Hidden sector}

In this subsection, we present the results for the hidden-sector scenario (outlined in Appendix \ref{subsec:PBH_hidden_sector}), in which PBH formation is driven by the vacuum energy difference between the true and false patches.
Figure~\ref{fig:GW_multiple_masses} shows the GW spectra of a purely hidden sector for different mass choices and hidden-sector scales $T_0$, while keeping the same vacuum energy density. The red, blue, and green curves correspond to $T_0 = 20~\text{MeV}$, whereas the black, gray, and magenta curves correspond to $T_0 = 100~\text{MeV}$. As expected, the GW spectrum shifts toward higher frequencies for larger $T_0$. Throughout, we choose $\beta/H \sim \mathcal{O}(700)$, which is close to the minimal value allowed for the polynomial-like potentials considered here \footnote{$\frac{\beta}{H} = \frac{2\,S_3}{T}\,\frac{(\eta - \kappa)\,f_S'(\kappa)}{f_S(\kappa )} $ \cite{Dent:2024bhi}, and since $S_3/T \sim O(140)$, $7\lesssim\frac{2\,f_S'(\kappa)}{f_S(\kappa)}\lesssim 14$, thus overall $\beta/H > 700$, and increases with increasing $\eta$. }. We compare our predictions with the power-law integrated sensitivity (PLIS) curves at a signal-to-noise ratio (SNR) threshold of 1 for several proposed GW detectors: the space-based interferometers LISA~\cite{Caprini:2019egz}, BBO~\cite{Crowder:2005nr}, Ultimate-DECIGO~\cite{Braglia:2021fxn}, and $\mu$Ares~\cite{Sesana:2019vho}. For completeness, we also show estimates of astrophysical foregrounds from galactic~\cite{Robson:2018ifk} and extragalactic compact binaries~\cite{Farmer:2003pa}.

Figure~\ref{fig:PBH_multi} shows the PBH dark matter abundance as a function of mass for two different $T_0$ values, with three benchmark $M_{\rm PBH}$ values for each $T_0$. Masses in the asteroid range arise naturally for the lower hidden-sector scale $T_0 = 20~\text{MeV}$, while larger values of $T_0$ lead to greater production of PBHs relative to the observed dark matter abundance. For $T_0 = 100~\text{MeV}$, obtaining an allowed relic density ($f_{\rm PBH} < 1$) requires lighter PBHs, placing the viable region within the orange exclusion band below about $10^{18}~\text{g}$. This region is excluded by Hawking evaporation bounds derived from a combination of observations: the extragalactic gamma-ray background~\cite{Carr:2009jm}, the CMB~\cite{Poulin:2016anj,Clark:2016nst}, dwarf galaxy heating~\cite{Kim:2020ngi}, the EDGES 21cm signal~\cite{Clark:2018ghm}, Voyager $e^{\pm}$ data~\cite{Boudaud:2018hqb}, the 511~keV gamma-ray line~\cite{DeRocco:2019fjq,Laha:2019ssq}, and the diffuse MeV Galactic flux~\cite{Laha:2020ivk}. The asteroid-mass range, roughly $10^{18}\text{--}10^{22}\,\text{g}$, remains unconstrained. Note that the upper bound on the allowed PBH masses is set by $\overline{M}$, which can in principle be much larger but would lead to an overabundance. Consequently, for FOPTs driven by polynomial-like potentials, we are restricted to asteroid-mass PBHs, as these potentials impose a lower bound on $\beta/H$ of order $\mathcal{O}(700)$. Introducing a non-zero cubic term ($A \phi^3$) in the potential relaxes this constraint, allowing for smaller $\beta/H$ values and thus higher PBH masses without overproduction.


\subsection{Fermi-balls}
\begin{figure}[h!]
    \centering
    \begin{subfigure}[b]{0.48\linewidth}
        \centering
        \includegraphics[width=\linewidth]{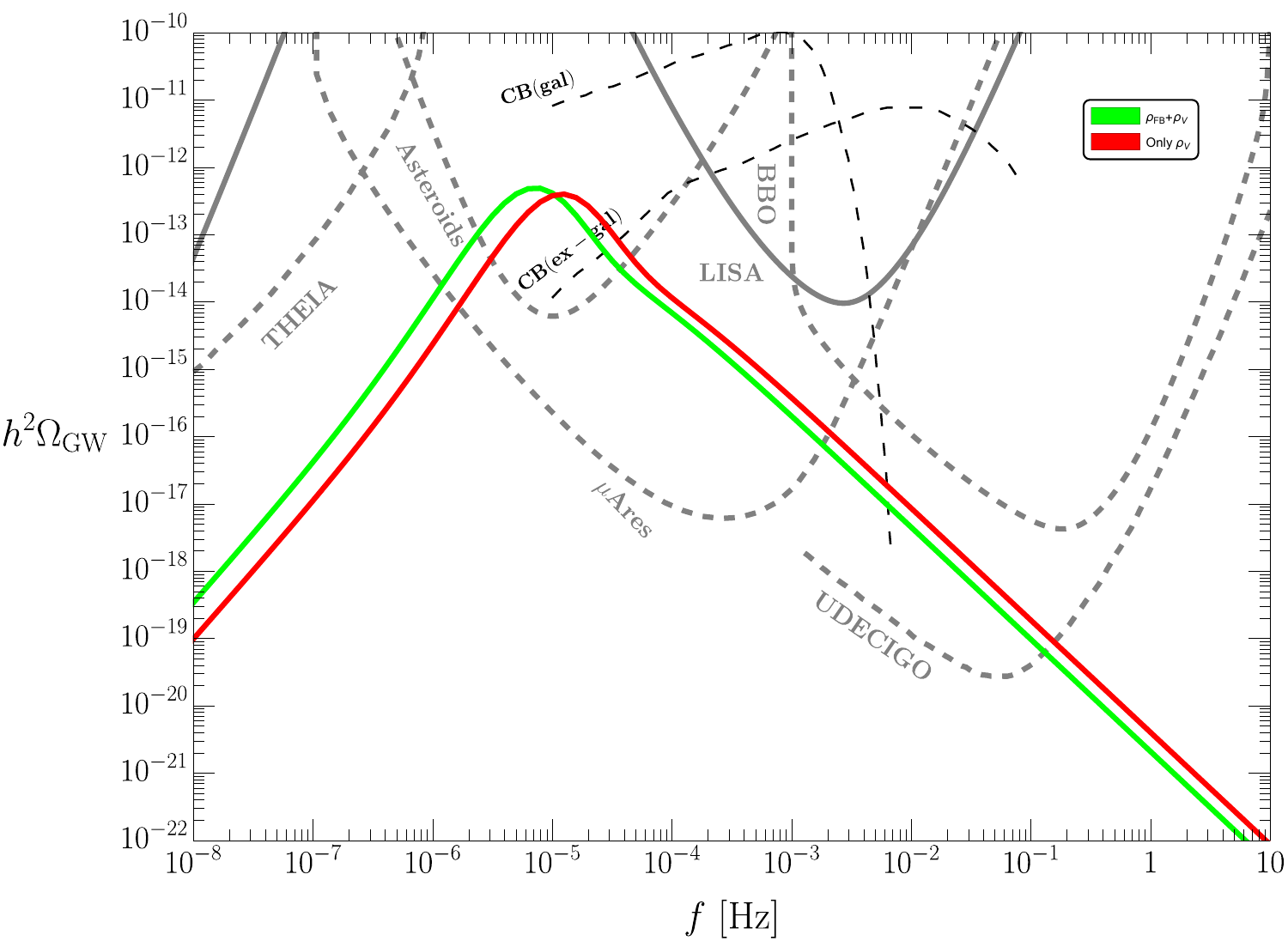}
    \end{subfigure}
    \hfill
    \begin{subfigure}[b]{0.48\linewidth}
        \centering
        \includegraphics[width=\linewidth]{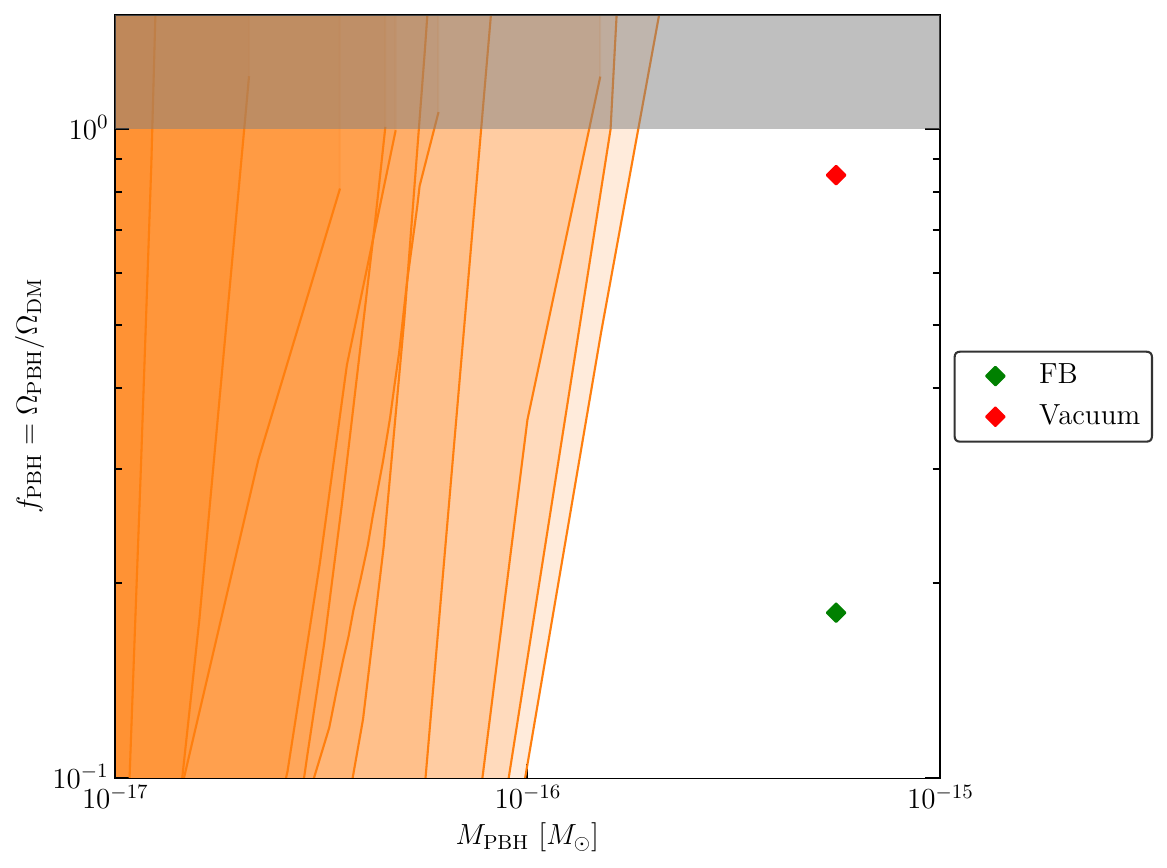}
    \end{subfigure}
    \caption{\justifying (Left) GW spectrum, $h^2\Omega_{\rm GW}$ vs $f[\rm Hz]$ for the Fermi-ball scenario. Here we have $(\rho_V+\rho_{FB}) / \rho_{rad}\sim0.085$, and $T_0 = 30\,\rm{MeV}$, $\xi = 1$, along with $\rho_{FB}\sim 3\,\rho_V$. (Right) For the Fermi-ball scenario, the red diamond denotes the pure-vacuum case, 
while the green diamond represents the FB+vacuum case, both corresponding 
to a PBH mass of $M_{\rm PBH} = 1.06 \times 10^{18}\,\text{g}$. }
    \label{fig:combined_FB}
\end{figure}

In this subsection, we present the results for the fermi-ball scenario (outlined in Section \ref{sec:PBH_FB}) and contrast them with the model scenario where the same amount of energy is stored in the vacuum.
In the left panel of Figure~\ref{fig:combined_FB} we compare the gravitational wave signal in these two scenarios. In the first (green curve), a Fermi-ball forms in addition to the hidden sector, while in the second (red curve) we consider only the vacuum energy density in the hidden sector, fixing its value $\rho_V/\rho_{rad}$ to equal the total energy w.r.to the radiation of the combined system ($(\rho_V + \rho_{FB})/\rho_{rad}$). This results in larger $\beta/H$ values for the pure vacuum scenario, as well as a slightly smaller nucleation temperature. The resulting GW spectra demonstrate that the purely vacuum-energy scenario is shifted towards higher frequency, which results from the increased $\beta/H$ values. The comparison of the GW peak amplitude ($\Omega_{GW,\text{peak}}$) shows that the pure vacuum case has a larger $\alpha$ than the Fermi-ball scenario, since part of the available energy in the latter goes into Fermi-ball formation rather than GW production. However, the pure vacuum case also features a larger $\beta/H$, which suppresses the GW amplitude. The interplay between these competing effects—enhancement from larger $\alpha$ and suppression from larger $\beta/H$—results in a GW signal with an amplitude comparable to that of the Fermi-ball scenario.
The right panel of Fig.~\ref{fig:combined_FB}  displays the PBH abundance scenarios -- the Fermi-ball plus vacuum (green diamond) and vacuum only (red diamond) cases. The vacuum only case corresponds to a higher value of $\beta/H$, which in turn leads to a larger PBH fraction $f_{\rm PBH}$. Overall, the combination of the results displayed in Figs.~\ref{fig:combined_FB} for the Fermi-ball plus vacuum and the vacuum only cases indicates that vacuum energy alone is more efficient for PBH formation and simultaneously yields stronger GW signals compared to scenarios where a portion of the energy is partitioned into Fermi-balls. 

These three plots illustrate the complementarity between the gravitational wave signal and the resulting primordial black hole abundance. A smaller value of $\beta/H$ enhances the GW amplitude while suppressing the PBH fraction of the dark matter. Conversely, increasing the hidden-sector temperature scale shifts the GW spectrum toward higher frequencies and simultaneously increases $f_{\rm PBH}$. The phase transition strength parameter $\alpha$ (see~\cite{Dent:2024bhi} for details of the calculation of $\alpha$) is directly related to the vacuum energy density $\rho_V$, which governs both PBH formation and the upper limit on the allowed PBH mass. Larger $\alpha$ leads to a higher amplitude of the associated GW signal. In these plots, we keep $\rho_V$ approximately constant, so that $\alpha$ scales primarily with $T_0$, highlighting the effect of the hidden-sector scale. Overall, these correlations establish an interesting link between observable GW signatures and PBH dark matter phenomenology.

The key distinction between the pressure-based criterion and the junction-condition approach lies in the definition of the characteristic radius: the turning-point radius $r_{\rm TP}$ versus the force-balance radius $r_{\rm FB}$. The turning-point condition arises from a dynamical equation derived from Einstein’s field equations, whereas the force-balance condition is a static relation obtained by requiring the net force to vanish. Once the system evolves past the turning point, it inevitably collapses to the Schwarzschild radius. We emphasize that this does not contradict thermodynamic arguments: if the force-balance condition is derived directly from Einstein’s equations rather than from minimizing a classical potential energy (which only identifies where the acceleration vanishes, without elucidating any information about the velocity), then the results should be essentially equivalent. In any case, the quantitative impact on the PBH abundance is minor, since Eq.~(\ref{eq:TPdep_fpbh}) is nearly constant as a function of $r$.
Also, we need to emphasize that we probe scenarios where the $\rho_{FB}/\rho_V \sim O(1)$, in order to compare it with scenario of only vacuum without overproducing dark matter.
\begin{figure}[h!]
    \centering
    \begin{subfigure}[b]{0.48\linewidth}
        \centering
        \includegraphics[width=\linewidth]{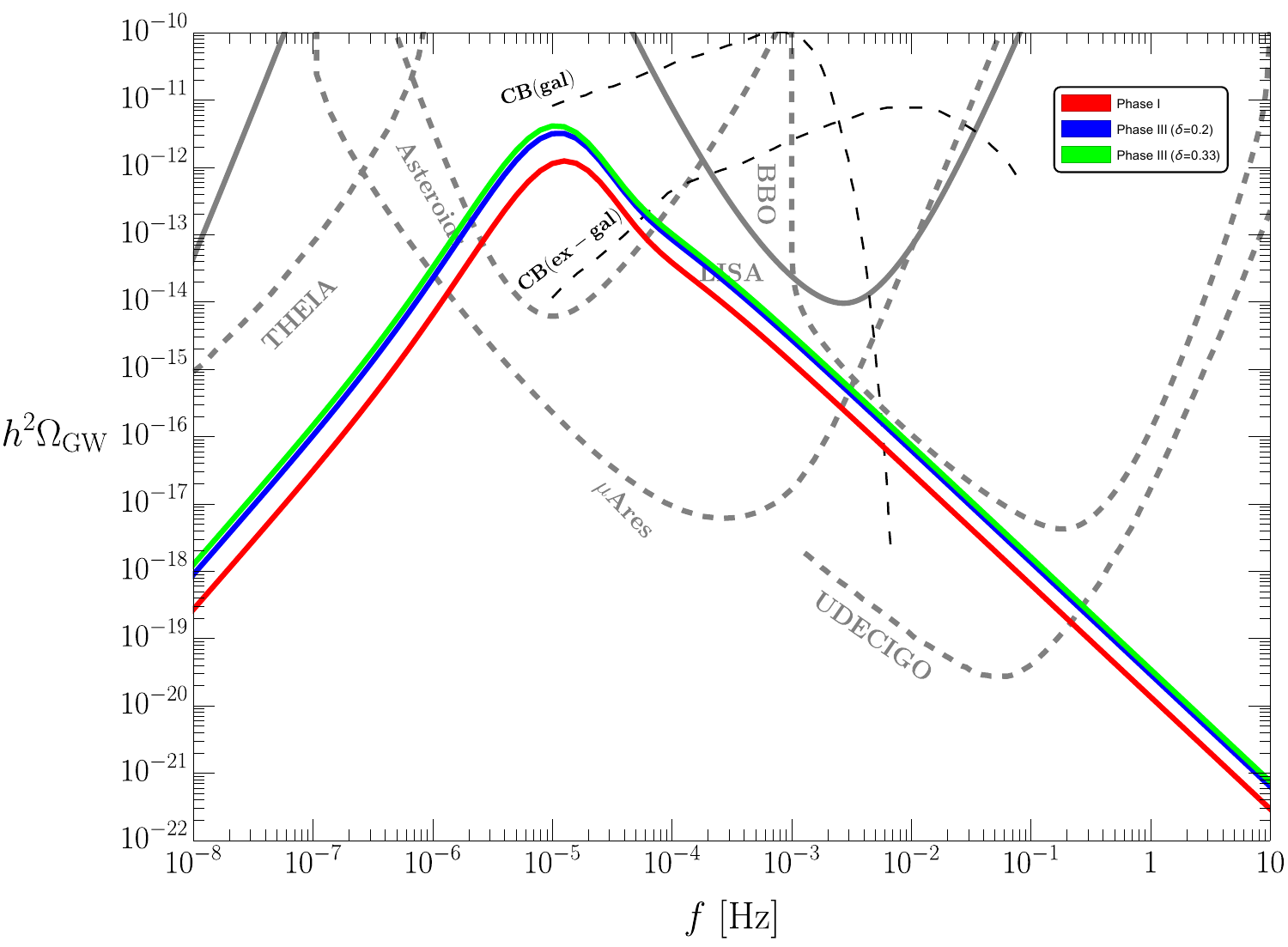}
        \label{fig:GW_multi_trans}
    \end{subfigure}
    \hfill
    \begin{subfigure}[b]{0.48\linewidth}
        \centering
        \includegraphics[width=\linewidth]{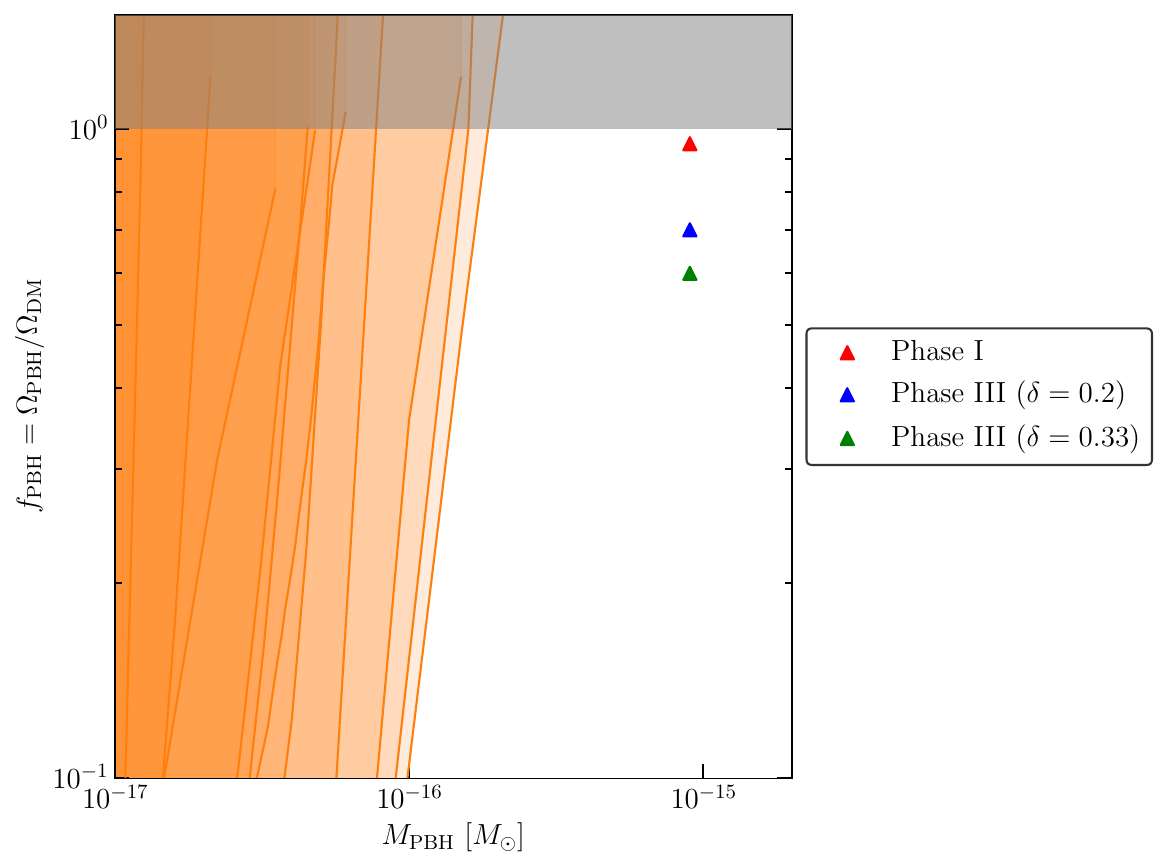}
        \label{fig:PBH_multi_tr}
    \end{subfigure}
    \caption{\justifying (Left) The GW spectrum corresponding to the multiple phases : Phase 1 (red), $\xi = 1$, Phase 3 for for a thermal kick of $\delta =0.2$ (blue) and Phase 3 for a thermal kick of $\delta =0.33$  (green) where $\xi_f = \xi(1+\delta)$ and hidden sector scale $T_0 = 50 \,\rm{MeV}$ and $\rho_V/\rho_{rad} \sim 0.08$ for the three scenarios. (Right) The PBHs produced in the three scenarios, where we match the coloring scheme to the GW signals. The masses of the PBH is $M = 1.8\times10^{18}\,\rm{g}$.}
    \label{fig:combined_multi_trans}
\end{figure}

\subsection{Multiple phase transition}
In the final set of plots displayed in Figure~\ref{fig:combined_multi_trans}, we examine scenarios with multiple phase transitions, as outlined in Section~[\ref{sec:PBH_multi}] for a given set of parameters described in the caption. The left panel of shows the GW spectra for two representative transitions: the blue curve corresponds to the initial Phase~1 transition, while the red curve corresponds to the third, Phase~3. Although the underlying microphysics of the two phases are nearly identical, differing only through the cosmological background in which they occur \cite{Dent:2024bhi}, the GW signal from Phase~3 is enhanced due to the transition taking place at a later epoch. The right panel displays the resulting $f_{\rm PBH}$ as a function of the PBH mass. A similar trend is observed as in the previous cases: transitions at lower temperature scales yield a relatively suppressed PBH abundance, with the scaling given by, 
\[
\frac{f_{\rm PBH,1}}{f_{\rm PBH,3}} \sim 
\left(\frac{g_S(T_{\rm TP})}{g_S\big(T_{\rm TP}/(1+\delta)\big)}\right)^{-1} \, (1+\delta)^3
\]
Since Phase~3 occurs after Phase~1, the thermal kick imparted to the hidden sector reduces the PBH abundance. 
This suppression is partially compensated by the ratio of entropy degrees of freedom, $g_s$, which decreases 
as the Universe cools. Moreover, the turning point temperatures are nearly equal, 
$T_{\rm TP,3} \simeq T_{\rm TP,1}$, and the dependence on $\beta/H$ remains mild, 
as both phases share similar nucleation dynamics. This picture is consistent with the expected gravitational 
wave signals: Phase~3 experiences less redshifting than Phase~1, leading to a stronger signal, 
as shown in Fig.~\ref{fig:combined_multi_trans}. Finally, increasing $\delta$ further softens the difference 
between the two phases, since the scaling always proceeds with powers of $(1+\delta)$, 
which remains close to unity unless $\delta = \mathcal{O}(1)$. Larger values would either dominate the 
total energy density of the Universe or lead to significant Hubble heating, both of which are excluded. By combining gravitational wave measurements with primordial black hole observations (from microlensing and Hawking radiation), one can determine the value of $\delta$.


\section{Conclusions}
\label{sec:conclusions}

Primordial black holes may play a crucial role in cosmology -- they could provide some or all of the dark matter, source gravitational waves, or play a role in early structure formation by seeding the formation of the supermassive black holes observed at redshifts $z\gtrsim 6$. One PBH formation avenue -- that they form at the endpoint of cosmic first-order phase transitions --  has increasingly become an area of interest as the possibility of correlated signatures from the GWs associated with the phase transition present an intriguing multi-messenger scenario. 

In this work we have revisited two theoretical descriptions of PBH formation pathways from a phase transition. These include: i) particle trapping in a false vacuum bubble, which has typically been described using a balance of pressures that initially stabilizes the false vacuum bubble before it collapses to a PBH, and ii) a vacuum energy dominated bubble which reaches a turnaround point, followed by a reduction in size until its bubble wall gradient energy induces black hole formation once the Schwarzschild criteria is achieved. We have extended the work of~\cite{Flores:2024lng} by applying the Israel junction conditions to analyze the PBH formation for each of these scenarios (particle trapping and gradient energy in the bubble walls) while allowing general energy densities within the false vacuum bubble. 

We have adopted a general polynomial potential for a hidden sector scalar field which undergoes the first-order phase transition. We calculated the GW spectrum for the case of a hidden sector with a transition temperature of $T_0\leq 100~\rm MeV$ along with the range of PBH masses $M_{\rm PBH}$ and dark matter mass fraction $f_{\rm PBH}$ of the PBH population produced from the phase transition. We found that PBHs can be produced within the asteroid mass window where the PBHs could form some or all of the dark matter, and that there is an upper limit to $M_{\rm PBH}$ (below the mass values covered by lensing observations) for viable values of $f_{\rm PBH}$ (that it be less than or equal to unity) due to the large parametric values that govern the transition rate. The correlated gravitational wave signatures from these transitions tend to peak at frequencies between the LISA band and frequencies to which pulsar timing arrays are sensitive. The peak frequency is determined by the temperature at which the transition occurs, but also feeds into $f_{\rm PBH}$, and consideration of temperatures that lead to frequencies in the LISA band would run afoul of the $f_{\rm PBH} \leq 1$ constraint.

We have shown that the peak of the GW signals is shifted towards higher frequency when the false vacuum bubble contains trapped particles -- as in the Fermi-ball scenario -- relative to the vacuum only situation, as the rate of transition increases when balancing energy compared to the Fermi-ball formation. Additionally, the $f_{\rm PBH}$ value for the Fermi-ball case is also smaller than the vacuum only case.

When considering a multiple phase transition scenario where the first phase transition is followed by a reheating scenario that produces two additional transitions, we find that the first transition produces GWs with a suppressed amplitude relative to the third phase transition, while the $f_{\rm PBH}$ values give the reverse scenario where the first transition produces roughly twice the relative abundance of the third phase. In these scenarios, one expects to obtain a distribution of PBH masses.



\section*{Acknowledgments} 

We thank Tao Xu, Jason Kumar, Tong Ou and Michael Baker for helpful comments. The work of BD and MR is supported by the U.S. Department of Energy Grant DE-SC0010813.  JBD acknowledges support from the National Science Foundation under grant
no. PHY-2412995. JBD thanks the Mitchell Institute at Texas A\&M University for its hospitality where part
of this work was completed.

\appendix
\section{Hidden sector model}
\label{app:hidden_sector_model}
We consider a polynomial potential for a hidden-sector scalar field $\phi$ that receives thermal corrections in the early universe,  
\[
V(\phi,T) \approx D(T^2-T_0^2)\phi^2 - E\,T \phi^3 + \frac{\lambda}{4} \phi^4,
\label{eq:V}
\]
where $D$, $E$, and $\lambda$ are model-dependent parameters. This form is widely studied and chosen here for its analytical tractability \cite{Adams:1993zs,Ellis:2020awk}, although our discussion applies more generally to models supporting a strong first-order phase transition (FOPT).

It is convenient to introduce the dimensionless parameter
\[
\eta \coloneqq \frac{2\lambda\,D}{E^2}, \qquad \eta > \tfrac{9}{4}.
\label{eq:eta}
\]
A useful dimensionless parameter entering the semi-analytical expressions for the Euclidean action, as well as the strength and rate of the phase transition, is  
\[
\kappa(T) = \frac{2\lambda D\,(T^2-T_0^2)}{E^2 T^2} 
= \eta\left(1-\frac{T_0^2}{T^2}\right), 
\qquad 0 \leq \kappa \leq 2.
\label{eq:kappa}
\]

At high temperatures ($T \gg T_0$), the potential has a single minimum at $\phi = 0$. As the universe cools below 
\[
T_1^2 = \frac{T_0^2}{1-\tfrac{9}{4\eta}},
\label{eq:T1}
\]
a second minimum develops. At the critical temperature $T_c$, this minimum becomes degenerate with the symmetric one:
\[
T_c^2 = \frac{T_0^2}{1-\tfrac{2}{\eta}}, 
\qquad \phi_c = \frac{2E}{\lambda}\,T_c.
\]
Finally, the zero-temperature vacuum expectation value (vev) is
\[
v^2 = \frac{2D}{\lambda}\,T_0^2.
\]

The vacuum energy, which vanishes at $T=0$, is given by
\[
\rho_V = \frac{\lambda v^4}{4}.
\label{eq:rhoV}
\]
In the thin-wall limit\footnote{The characteristic wall thickness is $l \approx \tfrac{2\sqrt{2}}{\lambda\,\phi_c}$} \cite{Flores:2024lng}, the surface tension can also be expressed in terms of the model parameters as  
\[
\sigma = \frac{\phi_c^3}{6}\sqrt{\frac{\lambda}{2}}.
\label{eq:sigma}
\]

\subsection{Israel Junction Condition \label{sec:Junction_conditin}}

The Israel junction condition\cite{Israel:1966rt} relates the jump in the extrinsic curvature
across a hypersurface $\Sigma$ to the surface stress--energy tensor
localized on it:
\begin{equation}
\big[ K_{ij} \big] - h_{ij}\,\big[ K \big] \;=\; - 8 \pi G \, S_{ij},
\end{equation}
where:
\begin{itemize}
    \item $h_{ij}$ is the induced metric on the hypersurface $\Sigma$, obtained by projecting the bulk metric $g_{\mu\nu}$ onto $\Sigma$,
    \item $K_{ij}$ is the extrinsic curvature of $\Sigma$, defined as 
    \(
      K_{ij} = h_i^{\;\mu} h_j^{\;\nu} \nabla_\mu n_\nu
    \),
    with $n_\mu$ the unit normal to $\Sigma$,
    \item $S_{ij}$ is the surface stress--energy tensor localized on $\Sigma$,
    \item $[X] \equiv X^{+} - X^{-}$ denotes the discontinuity of a quantity across $\Sigma$.
\end{itemize}

This condition provides the correct matching of two spacetimes at a thin shell
or domain wall and determines the dynamics of the surface layer.

\subsection{PBH Formation from phase transition in hidden sector}
\label{subsec:PBH_hidden_sector}
We will use the Israel-Junction conditions as the physical reasoning behind the possibility for the formation of PBHs from the phase transitions involving the polynomial like potentials. It is given as \cite{Blau:1986cw,Berezin:1987bc},
\[
M = \frac{4\,\pi}{3}\Delta\,\rho\,r^3 - \pi\frac{\sigma^2}{M_{pl}^2}
\,r^3+4\pi\sigma\,r^2\sqrt{1-\frac{\rho_{in}}{3\,M_{pl}^2}r^2+\dot{r}^2}
\]
where $\Delta\,\rho = \rho_{in} - \rho_{out} = \rho_V$ and $\dot{} \equiv d\tau$, $\tau$ being the physical time as measured along wall trajectory.
The above equation can be re-expressed in terms of dimensionless variables, 
\[
\left(\frac{d\,z}{d\tau'}\right)^2 + U(z) = E
\]
where, 
\[
U(z) = -\left(\frac{(1-z^3)}{z^2}\right)^2 - \frac{\gamma^2}{z}-\gamma^2(1-\frac{\gamma^2}{4})\,\left(\frac{\rho_{in}}{\rho_V}-1\right)\,z^2,\,
\]
and 
\[
E = -\gamma^2\left(1-\frac{\gamma^2}{4}\right)^{1/3}\,\left(\frac{\overline{M}}{M}\right)^{2/3},
\]
and \[
\gamma = 2\sin\theta,\,\quad \theta = \tan^{-1}\left(\frac{\sigma}{2\,M_{pl}\,\sqrt{\rho_V/3}}\right) \sim \frac{v}{M_{pl}},v\ll M_{pl}
\]
where $z^3 = \frac{4\pi\,\rho_V}{3\,M\,(1-\gamma^2/4)}\,r^3$ and $\overline{M} = \frac{4\pi\,M_{pl}^3}{\sqrt{\rho_V/3}} = \frac{8\pi\,M_{pl}^3}{v^2}\sqrt{\frac{3}{\lambda}}$ is the maximum mass allowed for the collapsing PBH. $\sigma$ and $\rho_V$ are defined in Eq.(\ref{eq:sigma}) and Eq.(\ref{eq:rhoV}) respectively.
Solution for the above equation yields the collapse time (when converted back to dimensionful quantities) and is given by,
\[
\tau'=\int \frac{dz'}{\sqrt{E-U(z')}}
\]
where $\tau' = \tau\,\frac{\sqrt{\rho_V/3}}{M_{pl}\,\gamma\,\sqrt{1-\gamma^2/4}}$.
The turning point, where $E = U(z)$ occurs, corresponds to the point beyond which the volume will collapse into a black hole. Thus we are interested in this point, which we define as $z_{TP}$ and the corresponding radius is defined as $r_{TP}$, with the equation given by \footnote{For most of the parameter space where $v\ll M_{pl}$, $z_{TP}\approx 1$.},
\[
\left(1-\frac{1}{z^3}\right)^2 + \frac{4\sin^2\theta}{z^2}\left(\,z^2\,\left(\frac{1}{z^3} + \cos^2\theta\frac{\rho_{rad}}{\rho_V}\right) - \cos^{2/3}\,\theta \left(\frac{\overline{M}}{M}\right)^{2/3}\right) = 0
\label{eq:zTP_vac}
\]

The Schwarzschild radius corresponds to $r_s = 2\,G\,M$, corresponding to 
\[
z_s = \frac{\gamma^2}{|E|} \approx \left(\frac{M}{\overline{M}}\right)^{2/3} \ll 1,
\]
and the corresponding horizon radius, $r = H_V^{-1}$, corresponding to 
\[
z_H =\frac{\sqrt{|E|}}{\gamma(1-\gamma^2/4)^{1/2}} = \left(\frac{\overline{M}}{M}\right)^{1/3} \gg 1
\]
\subsubsection{Conditions for PBH formation}
The following conditions must be satisfied in order to form a black hole \cite{Flores:2024lng},
\begin{itemize}
    \item The collapse time for the PBH should be much smaller than the corresponding Hubble time, i.e.,
    \begin{align}
            & \frac{\tau}{H^{-1}}\ll 1 \\
            \implies &\gamma\sqrt{1-\frac{\gamma^2}{4}}\, \tau'  \ll \frac{\sqrt{\rho_V/3}}{H\,M_{pl}}
    \end{align}
    \item We also want the collapse to slower than the nucleation time, since we need the true vacuum bubbles to form and trap false vacua patches. 
    \item The size of domain wall should be smaller than the Scharzchild radius, i.e., 
    \begin{align}
     & l \leq r_s \\
     \implies & \frac{\lambda\,\phi_c\,M}{8\pi\sqrt{2}M_{pl}^2} \geq 1
     \end{align}
\end{itemize}

\section{Comment on force balancing criteria for FB collapse \label{sec:App_FB}}

The general criterion for Fermi--ball collapse, following Refs.~\cite{Hong:2020est, Marfatia:2021hcp,Marfatia:2024cac}, is defined at the radius where  
\begin{equation}
\frac{\partial \Delta F}{\partial R} = 0 \, .
\end{equation}
where $\Delta$ is the difference between the false and true vacuum and we have \cite{Marfatia:2021hcp},  
\begin{equation}
\Delta F = \left(f_Q(r,T) + \rho_V \right)\frac{4\pi r^3}{3}, 
\qquad 
f_Q(r,T) = \frac{a}{r^4} + \frac{d}{r^2}-\frac{b}{r^6},
\end{equation}
Applying the condition yields\footnote{Including the surface term will yield : $r^6+2r^5\frac{\sigma_0}{\rho_V}+\frac{d}{3\rho_V}\,r^4-\frac{a}{3\,\rho_V}\,r^2+\frac{b}{\rho_V}=0$. The surface term is extremely suppressed and hence would not change the solution.}, 
\begin{equation}
0 = u^3 + \frac{d}{3c}u^2 - \frac{a}{3c}\,u + \frac{b}{c},
\end{equation}
where we have introduced $u = r^2$.  

To remove explicit dependence on the dimensionful variable $r$, we introduce a dimensionless coordinate $z$, defined earlier in Eq.(\ref{eq:dimless_vars}). This allows us to express the equation purely in terms of model parameters, extracting out the dependence on $\{Q, M\}$. Defining  
\begin{equation}
\tilde{z} = \left(\frac{M}{\rho_V^{1/4}Q}\right)^{2/3} z^2,
\end{equation}
the equation reduces to a cubic form\footnote{This equation is functionally different from the Eq.(\ref{eq:z_TP_trunc}) obtained from the junction conditions, and there is no \textit{apriori} reason that the solutions should match.}, 
\begin{equation}
\tilde{z}^3 + \tilde{a}_2 \tilde{z}^2 - \tilde{a}_1 \tilde{z} + \tilde{a}_0 = 0,
\end{equation}
with coefficients  
\begin{align}
    \tilde{a}_1 &= \frac{1}{4 \cos^{2}\theta}\,\left(\frac{3 \pi^2}{\cos^{2}\theta}\right)^{1/3}, \\
    \tilde{a}_0 &= \frac{g_\chi^2 L_\phi^2 \sqrt{\rho_V}}{2 \cos^4\theta}, \\
    \tilde{a}_2 &= \frac{T^2}{2 \sqrt{\rho_V}} \left(\frac{\pi^2 \cos\theta}{9} \right)^{2/3}.
\end{align}  

The solution of the cubic is given by  
\begin{equation}
\tilde{z} = \frac{1}{3} \left( 2 \sqrt{\tilde{a}_2^2 + 3 \tilde{a}_1}\,
\cos \!\left[\tfrac{1}{3} \cos^{-1}\!\left(\frac{-2 \tilde{a}_2^3 - 9 \tilde{a}_1 \tilde{a}_2 - 27 \tilde{a}_0}{2 (\tilde{a}_2^2 + 3 \tilde{a}_1)^{3/2}}\right)\right]
- \tilde{a}_2 \right),
\end{equation}
provided that,  
\begin{equation}
\frac{2 \tilde{a}_2^3 + 9 \tilde{a}_1 \tilde{a}_2 + 27 \tilde{a}_0}{2 \, (\tilde{a}_2^2 + 3 \tilde{a}_1)^{3/2}} \leq 1.
\label{eq:r_FB_criteria}
\end{equation}
The corresponding energy density, expressed in terms of $\tilde{z}$, is  
\begin{equation}
\frac{\rho_{FB}(\tilde{z})}{\rho_V} = \frac{3 \tilde{a}_1}{\tilde{z}^2} - \frac{3 \tilde{a}_2}{\tilde{z}} - \frac{\tilde{a}_0}{\tilde{z}^3}.
\end{equation}
Also, the condition  
\begin{equation}
\frac{\rho_{FB}(z)}{\rho_V} > 1
\end{equation}
is equivalent to  
\begin{equation}
\frac{\rho_{FB}(\tilde{z})}{\rho_V} > 1,
\end{equation}
and is therefore independent of the specific values of $\{Q, M\}$, depending only on the model parameters. For the physical solutions of $z_{FB}$, this criterion is indeed satisfied.  

\section{Trapped particles leading to PBH }
\label{app:trapping}
In this section, we briefly analyze the scenario envisioned in \cite{Baker:2021nyl}, where a particle is effectively massless inside the false vacuum but acquires a large mass in the true vacuum, $m_\chi = g\,v(T) > T$. As a result, it becomes trapped inside the false vacuum.  

As an illustrative example, consider a fermionic field $\chi$ coupled to a scalar field $\phi$ undergoing a first-order phase transition,
\begin{equation}
\mathcal{L} \supset g\,\phi\,\bar{\chi}\chi + V(\phi,T).
\end{equation}
In this scenario, $\chi$ remains in thermal equilibrium with $\phi$ in the false vacuum. Since the field is much heavier than the temperature in the true vacuum, it becomes non-relativistic and carries negligible energy density. Thus, overall the net energy difference is given by,
\begin{equation}
\Delta \rho_\chi \approx \rho_{\chi,f} = \frac{7\pi^2}{240}\,g_\chi^*\,T^4,
\end{equation}
where $g_\chi^* = 4$ accounts for the relativistic degrees of freedom of $\chi$ inside the false vacuum.  

Thus, the total energy difference between the vacua is
\begin{equation}
\Delta \rho = \Delta \rho_\chi + \rho_V.
\end{equation}
For the trapped fermion contribution to significantly affect the turning point radius, we require
\begin{equation}
\Delta \rho_\chi > \rho_V 
\;\;\;\implies\;\;\;
\frac{7\pi^2}{240}\,g_\chi^* T^4 > \frac{\lambda\,v^4}{4}
\;\;\;\implies\;\;\;
\frac{7\pi^2\,\lambda}{60\,D^2}\left(\frac{1}{1-\kappa/\eta}\right)^2 > 1.
\end{equation}

The Israel junction condition is then modified to
\begin{equation}
M = \frac{4\pi}{3}\,\Delta\rho\,r^3 - \pi\,\frac{\sigma^2}{M_{pl}^2}\,r^3
+ 4\pi\sigma\,r^2\sqrt{1-\frac{\rho_{\rm in}}{3M_{pl}^2}r^2+\dot{r}^2},
\end{equation}
where the calculation proceeds as in the standard case, with the replacement $\rho_V \rightarrow \rho_V + \rho_\chi$.

\begin{figure}[H]
    \centering
    \begin{subfigure}[b]{0.48\linewidth}
        \centering
        \includegraphics[width=\linewidth]{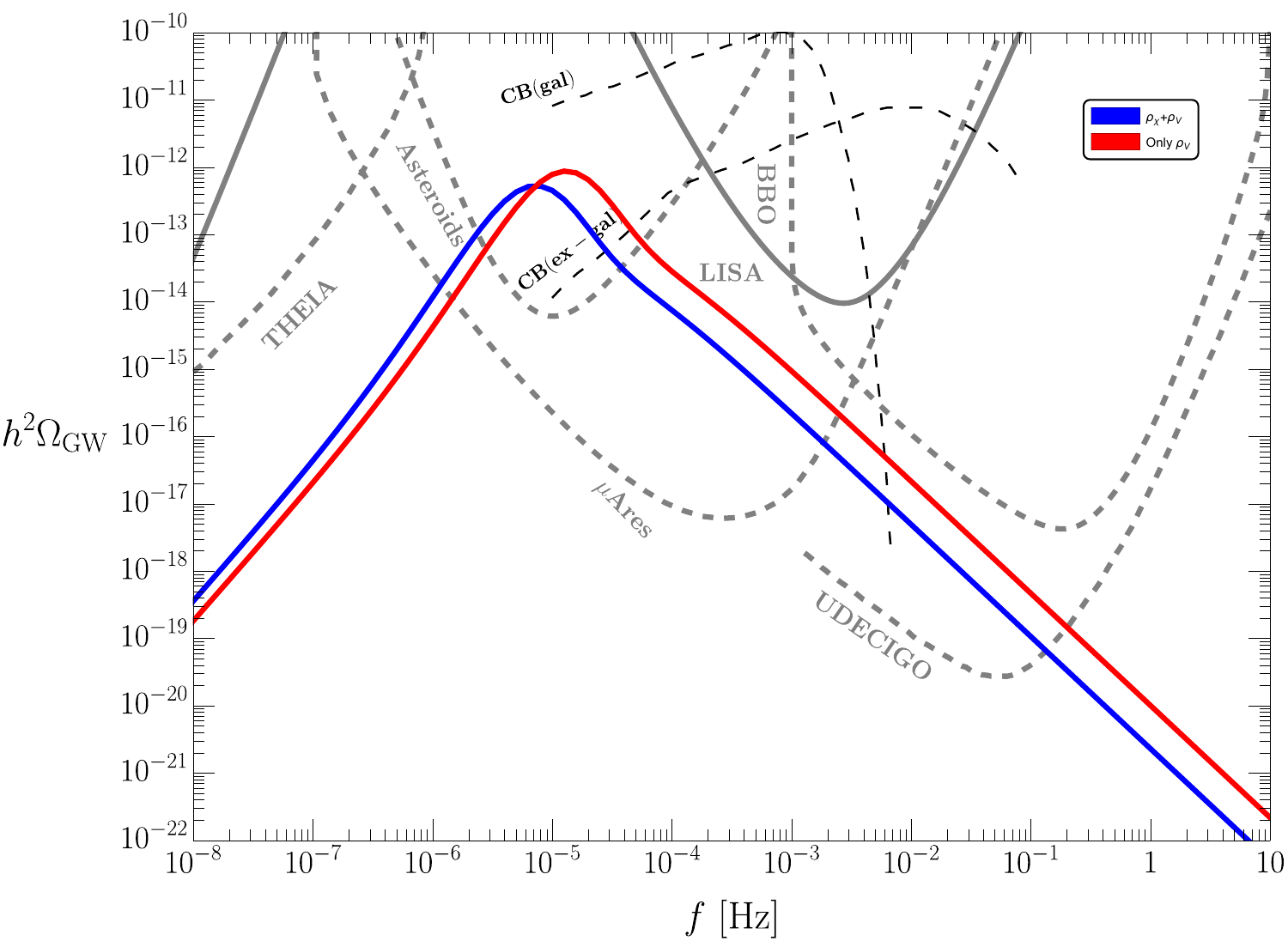}
    \end{subfigure}
    \hfill
    \begin{subfigure}[b]{0.48\linewidth}
        \centering
        \includegraphics[width=\linewidth]{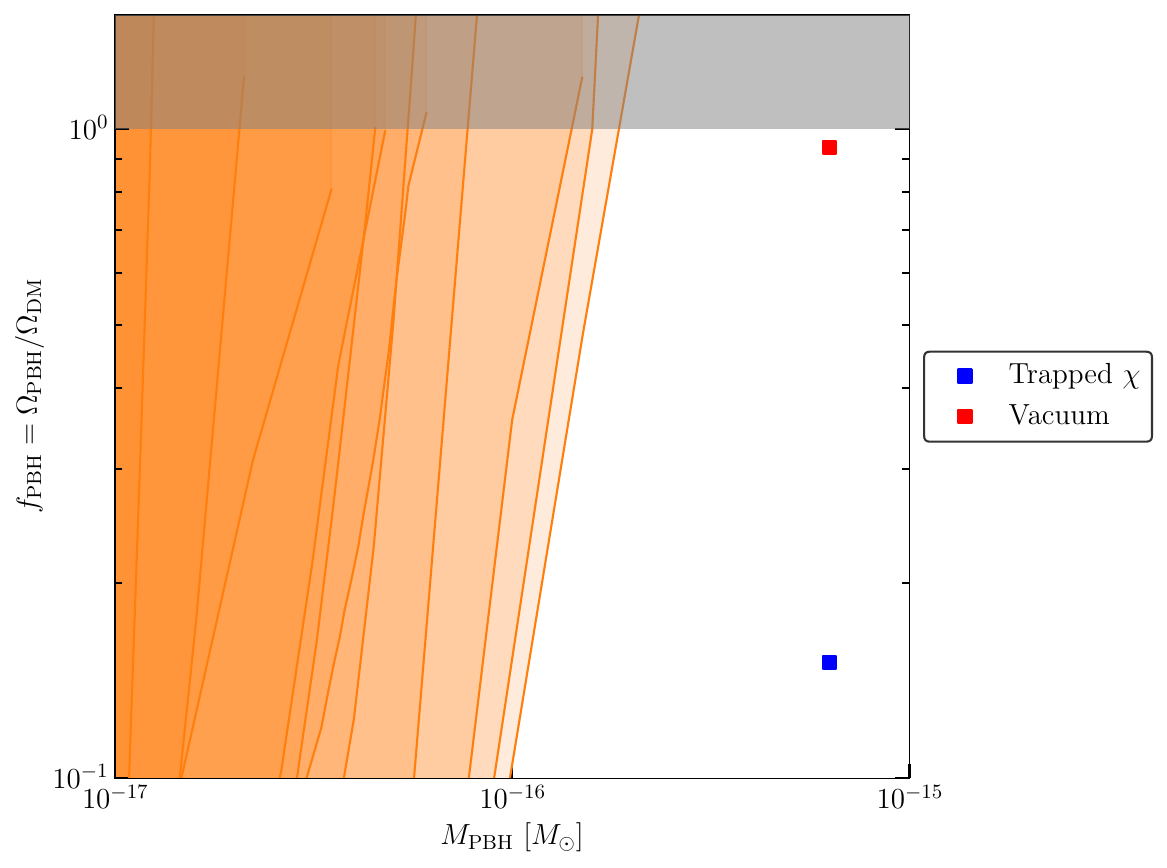}
    \end{subfigure}
    \caption{(Left) The GW spectrum corresponding to fermion trapping (blue) vs pure vacuum (red), where $\xi = 1$, hidden sector scale $T_0 = 30 \,\rm{MeV}$ and $(\rho_\chi+\rho_V)/\rho_{rad} \sim 0.2$ and $\rho_\chi/\rho_V \sim 9$. (Right) The PBH abundance  produced from the two scenarios where the red square corresponds to vacuum only and blue square corresponds to trapped fermion in addition to vacuum. The masses of the PBH is $M = 1.3\times10^{18}\,\rm{g}$.}
    \label{fig:trap}
\end{figure}

In terms of our dimensionless variables, the evolution equation becomes
\begin{equation}
\left(\frac{dz}{d\tau'}\right)^2 + U(z) = E,
\end{equation}
with
\begin{align}
U(z) &= -\left(\frac{1-z^3}{z^2}\right)^2 - \frac{4\sin^2\theta}{z}\left(1+
\cos^2\theta
\left(\frac{\rho_{\rm in}}{\rho_V+\rho_\chi}-1\right)z^3\right), \\
E &= -4\sin^2\theta\,
\left(\frac{\cos\theta\,\overline{M}}{M}\right)^{2/3},
\end{align}
where
\begin{equation}
z^3 = \frac{4\pi\,(\rho_V+\rho_\chi)}{3M\,\cos^2\theta}\,r^3,
\qquad
\overline{M} = \frac{4\pi\,M_{pl}^3}{\sqrt{(\rho_V+\rho_\chi)/3}}
\end{equation}
is the maximum mass allowed for the collapsing PBH, and 
\begin{equation}
\theta = \tan^{-1}\left(\frac{T_0}{M_{pl}}
\frac{8\sqrt{5}\,\bigl(\tfrac{D}{\eta-2}\bigr)^{3/2}(\eta-\kappa)}
{\sqrt{\lambda\left(60 D^2(\eta-\kappa)^2 + 7\pi^2\eta^2\lambda\right)}}\right) \ll 1 \,\, \forall \,\,T_0\ll M_{pl}.
\end{equation}
The turning point is defined where $U(z) = E$, and leads to the following equation,
\[
\left(1-\frac{1}{z^3}\right)^2 + \frac{4\sin^2\theta}{z^2}\left(\,z^2\,\left(\frac{1}{z^3} + \cos^2\theta\frac{\rho_{rad}}{\rho_V+\rho_\chi}\right) - \cos^{2/3}\,\theta \left(\frac{\overline{M}}{M}\right)^{2/3}\right) = 0
\label{eq:zTP_trap}
\]
The resulting solutions are still located around $z_{TP} \simeq 1$. Importantly, this mechanism enlarges the parameter space for collapse: even in regions where the vacuum energy alone would not trigger PBH formation, the presence of a trapped fermion component can be sufficient to induce collapse. However, the trade-off appears in the gravitational-wave signal: for the same total energy density $\rho_V + \rho_\chi$, the strength of the transition is reduced compared to the pure vacuum case, since part of the energy is carried by fermionic degrees of freedom that do not contribute directly to the GW strength parameter $\alpha$, as shown in Figure(\ref{fig:trap}). 

In the plots below, we first consider the case of trapped fermions, and ensure that most of the energy inside the false patch is coming from the trapped fermion, where we choose $\rho_\chi/\rho_V\sim 10$. In the other scenario, we consider the case of pure vacuum, where we ensure that it has the same energy as the trapped fermion scenario. Overall, we verify the hierarchy in the GW signal, where the pure vacuum give rise to stronger GW amplitude. In order to have the same energy density, it requires larger $\beta/H$ value, thus yielding larger PBH abundance as well, as shown in the right panel of Figure \ref{fig:trap}. As expected, the trapped fermion scenario is similar to the case of fermi-balls in terms of observed GW and the PBH abundance.

\section{Comment on local overdensity criteria for PBH formation}
\label{app:overdensity}

In a wide range of contemporary literature \cite{Carr:1975qj, Khlopov:1980mg, Musco:2004ak, Carr:2021bzv}, the formation of primordial black holes (PBHs) has often been characterized using a local overdensity criterion. Defining the overdensity as
\begin{equation}
\delta(t) = \frac{\rho_{\rm in}(t)}{\rho_{\rm out}(t)} - 1 = 
\frac{\Delta \rho(t)/\rho_{\rm in}(t)}{1-\Delta \rho(t)/\rho_{\rm in}(t)},
\end{equation}
for $\delta \ll 1$ one has $\delta \approx \Delta \rho(t)/\rho_{\rm in}(t)$. PBH collapse is expected to occur when the overdensity exceeds a critical threshold, commonly taken to be $\delta_c \sim 0.45$ \cite{Musco:2018rwt}, i.e.,
\begin{equation}
\delta(t_{\rm PBH}) \sim 0.45 \quad \implies \quad \frac{\Delta \rho(t_{\rm PBH})}{\rho_{\rm in}(t_{\rm PBH})} \sim 0.31.
\end{equation}

Physically, this condition selects a temperature at which the collapse can occur. For a polynomial-like potential, with temperature $T$ and a Standard Model (SM) bath at $T_{\rm SM} = T/\xi$, one can relate this condition to the model parameters. We have
\begin{equation}
\Delta \rho = \frac{\lambda\,v^4}{4}, \qquad v^2 = \frac{2\lambda}{D} T_0^2,
\end{equation}
and during the radiation-dominated era,
\begin{equation}
\rho_{\rm rad} = \frac{\pi^2}{30}\left(\xi^{-4} g_{\rm SM}^* + g_h^*\right) T^4 \equiv \frac{\pi^2}{30} g_{\rm eff}^* T^4.
\end{equation}

Using the parametrization $\kappa_{\rm BH} = \eta \left(1 - T_0^2 / T_{\rm BH}^2\right)$, we find
\begin{equation}
\kappa_{\rm BH} \approx \eta \left(1 - \frac{1.54 \sqrt{g_{\rm eff}^*\,\lambda}}{D}\right).
\end{equation}

By selecting appropriate points in parameter space, one can estimate the temperature at which the overdensity criterion is satisfied. Since $g_{\rm eff}^* \sim \mathcal{O}(10-100)$ even for $\xi = 1$, one requires $\sqrt{\lambda}/D \lesssim 0.1$ to satisfy $\kappa_{\rm BH} < 2$. For $\xi < 1$, this condition becomes even more restrictive, indicating that the overdensity criterion can impose stronger constraints than those derived from the Israel junction conditions.

The same analysis can be expressed in terms of the nucleation (or critical) time $t_N$ with $\kappa_N = \eta \left(1 - T_0^2 / T_N^2\right)$:
\begin{equation}
t_{\rm BH} \approx 1.59\, t_N \left(1 - \frac{\kappa_N}{\eta}\right) \frac{\sqrt{\lambda\,g_{\rm eff}^*}}{D} 
\approx 1.59\, t_c \left(1 - \frac{2}{\eta}\right) \frac{\sqrt{\lambda\,g_{\rm eff}^*}}{D}.
\end{equation}
Clearly, the time required to reach the overdensity threshold occurs significantly later than the turning point radius, which leads to a lower PBH abundance because the false-vacuum regions have a reduced probability of survival. In scenarios involving Fermi-balls or Q-balls, one can replace $\Delta \rho \rightarrow \rho_V + \rho_{QB}$, which facilitates satisfying the overdensity criterion.

\section{Gravitational waves parametrization}
\label{app:GW_param}
Gravitational waves originating from first-order phase transitions in a hidden sector can be generated through three primary processes: bubble wall collisions, sound waves in the plasma, and magneto-hydrodynamic (MHD) turbulence. Independently of the production channel, the stochastic GW background is described by the differential GW energy density parameter~\cite{Huber:2008hg,Caprini:2009yp}, defined as
\begin{align}\label{eq:spectrum}
\Omega_\text{GW}(f) \equiv
\frac{1}{\rho_c} \, \frac{d\rho_\text{GW}(f)}{d\log f},
\end{align}
with $\rho_c = 3H^2 / (8\pi G_N)$ denoting the critical energy density.

The cosmic expansion redshifts both the frequency and energy density of gravitational waves~\cite{Breitbach:2018ddu}, yielding
\begin{align}
\label{eq:full-redshift}
\Omega_\text{GW}^0(f) &= \mathcal{R}\,\Omega_\text{GW}\left(\tfrac{a_0}{a}f\right),
\end{align}
with
\begin{equation}
\mathcal{R} \equiv \left(\tfrac{a}{a_0}\right)^{4}\left(\tfrac{H}{H_0}\right)^{2}
\simeq 2.473 \times 10^{-5} \, h^{-2}
\left(\tfrac{g_s^\text{EQ}}{g_s}\right)^{4/3}
\left(\tfrac{g_\rho}{2}\right).
\end{equation}
Here $\Omega_\text{GW}^0(f)$ and $\Omega_\text{GW}(f)$ refer to the present and emission spectra, while $a_0$ ($a$) denotes the scale factor today (at nucleation).
For runaway transitions, the GW spectrum at emission admits a semi-analytical parametrization~\cite{Huber:2008hg,Caprini:2009yp,Hindmarsh:2015qta},
\begin{align}
\Omega_\text{GW}(f) \simeq
\sum_i \mathcal N_i, \Delta_i(v_w)
\left(\frac{\kappa_i \alpha}{1+\alpha}\right)^{p_i}
\left(\frac{H}{\beta}\right)^{q_i}
s_i(f/f_{p,i}),
\label{eq:OmegaGW-emission}
\end{align}
where $i \in \{\text{BW}, \text{SW}, \text{turb}\}$ labels the contributions from bubble walls, sound waves, and turbulence. From Eq.~\eqref{eq:OmegaGW-emission}, the GW spectrum is governed by the transition strength $\alpha$, the rate of transition $\beta/H$, the wall velocity $v_w$, and the efficiency factors $\kappa_{\rm BW}$ (energy in bubble walls) and $\kappa_{\rm SW}$ (energy in bulk plasma motion). $f_p$ corresponds to the peak frequency, while $s(x)$ are the spectral shape functions. The sound wave efficiency factor is given by~\cite{Ellis:2019oqb},
\[
\kappa_{\rm SW} = (1-\kappa_{BW})\,\frac{\alpha_{eff}}{0.73+0.083\sqrt{\alpha_{eff}}+\alpha_{eff}},\quad \alpha_{eff} = \alpha_h(1-\kappa_{\rm BW})
\]
Following  Refs.~\cite{Caprini:2019egz,Huber:2008hg}, we have the normalization factors given  by $(N_{\rm BW},N_{\rm SW},N_{\rm turb}) = (1, 0.159, 20.1)$.  
The exponents are given by $(p_{\rm BW},p_{\rm SW},p_{\rm turb}) = (2,2,3/2)$ and $(q_{\rm BW},q_{\rm SW}, q_{\rm turb}) = (2,1,1)$. The velocity factors, spectral shape functions, and corresponding peak frequencies are taken to be \cite{Breitbach:2018ddu}
\begin{align}
    \label{eq:spectrum}
    & \Delta_{\rm BW}=\frac{0.11 v^{3}_{\rm w}}{(0.42 + v^{2}_{\rm w})},\quad f_{\rm p, BW} = \frac{0.62\beta}{1.8-0.1v_{\rm w}+v_{\rm w}^2},\quad s_{\rm BW}(x) = \frac{3.8 \, x^{2.8}}{1+2.8 \, x^{3.8}},   \nonumber \\ & \Delta_{\rm SW}=v_{\rm w}\min\,(1,H_*\tau_\mathrm{sh}),\qquad  f_{\rm p, SW} = \frac{2\beta}{\sqrt{3}v_{\rm w}},\qquad s_{\rm SW}(x) =x^{3}\left(\frac{7}{4+3 \, x^{2}}\right)^{7/2},
    \nonumber \\ & \Delta_{\rm turb}=v_{\rm w},\qquad  f_{\rm p, turb} = \frac{3.5\beta}{2v_{\rm w}},\qquad s_{\rm turb}(x) =\frac{x^{3}}{(1+x)^{11/3}(1+8\pi\,x \frac{f_{\rm p, turb}}{H})}.
\end{align}

For non-runaway transitions, the bubble collision contribution is negligible and subject to large uncertainties, so we omit it~\cite{Basler:2024aaf}. The GW spectrum then arises from sound waves and turbulence, with the same spectral functions as in the runaway case. The distinction lies in the modified sound-wave efficiency factor,\cite{Espinosa:2010hh,Basler:2024aaf}
\begin{align}
 	\kappa_\mathrm{SW}&=
	 \begin{cases}
      \frac{c_s^{11 / 5} \kappa_A \kappa_B}{\left(c_s^{11 / 5}-v_w^{11 / 5}\right) \kappa_B+v_w c_s^{6 / 5} \kappa_A}, & \text{if}\ v_w<c_s \\
      \kappa_B+\left(v_w-c_s\right) \delta \kappa+\frac{\left(v_w-c_s\right)^3}{\left(v_J-c_s\right)^3}\left[\kappa_C-\kappa_B-\left(v_J-c_s\right) \delta \kappa\right], & \text{if}\ c_s<v_w<v_J \\
      \frac{\left(v_J-1\right)^3 v_J^{5 / 2} v_w^{-5 / 2} \kappa_C \kappa_D}{\left[\left(v_J-1\right)^3-\left(v_w-1\right)^3\right] v_J^{5 / 2} \kappa_C+\left(v_w-1\right)^3 \kappa_D}, & \text{if}\ v_J<v_w 
	\end{cases}
 \label{eq: eff_factor_SW}
\end{align}
with
\begin{align}
	\kappa_A &\simeq v_w^{6 / 5} \frac{6.9 \,\alpha_h}{1.36-0.037
                   \sqrt{\alpha_h}+\alpha_h}, &  \kappa_B \simeq
  &\frac{\alpha_h^{2 / 5}}{0.017+\left(0.997+\alpha_h\right)^{2 / 5}},
    \nonumber \\
	\kappa_C &\simeq
                   \frac{\sqrt{\alpha_h}}{0.135+\sqrt{0.98+\alpha_h}}, &
                                                                     \kappa_D
  &\simeq \frac{\alpha_h}{0.73+0.083 \sqrt{\alpha_h}+\alpha_h}, \nonumber\\
	\delta \kappa &\simeq -0.9 \log \frac{\sqrt{\alpha_h}}{1+\sqrt{\alpha_h}}.
\end{align}

\bibliography{main}
\end{document}